\documentclass[12pt,letterpaper]{article}
\usepackage{epsfig,rotating,setspace,latexsym,amsmath,epsf,amssymb,amsfonts,bm,theorem,cite,
caption,subcaption,enumerate,longtable,accents,siunitx,booktabs,adjustbox}
\usepackage{algorithm,algorithmic,graphicx,epsf,authblk,epstopdf,url,color,multirow}
\usepackage{mathtools,comment,tabularx}
\usepackage{comment}

\newtheorem{theorem}{Theorem}

\newtheorem{remark}{Remark}
\newtheorem{lemma}{Lemma}

\newenvironment{Proof}[1]{\medskip\par\noindent{\bf Proof:\,}\,#1}{{\mbox{\,$\blacksquare$}\par}}

\newcolumntype{Y}{>{\centering\arraybackslash}X}
\allowdisplaybreaks

\begin{document}

\title{Network Connectivity--Information Freshness Tradeoff in Information Dissemination Over Networks\thanks{This work was presented in parts at Allerton Conference, Sept.~2023 and IEEE MILCOM, Oct.~2023.}}

\author{Arunabh Srivastava \qquad Sennur Ulukus\\
        \normalsize Department of Electrical and Computer Engineering\\
        \normalsize University of Maryland, College Park, MD 20742\\
        \normalsize  \emph{arunabh@umd.edu} \qquad \emph{ulukus@umd.edu}}

\date{}

\maketitle

\vspace*{-0.8cm}

\begin{abstract}
    We consider a gossip network consisting of a source generating updates and $n$ nodes connected according to a given graph structure. The source keeps updates of a process, that might be generated or observed, and shares them with the gossiping network. The nodes in the network communicate with their neighbors and disseminate these version updates using a push-style gossip strategy. We use the version age metric to quantify the timeliness of information at the nodes. We first find an upper bound for the average version age for a set of nodes in a general network. Using this, we find the average version age scaling of a node in several network graph structures, such as two-dimensional grids, generalized rings and hyper-cubes. Prior to our work, it was known that when $n$ nodes are connected on a ring the version age scales as $O(n^{\frac{1}{2}})$, and when they are connected on a fully-connected graph the version age scales as $O(\log n)$. Ours is the first work to show an age scaling result for a connectivity structure other than the ring and the fully-connected network, which constitute the two extremes of network connectivity. Our work helps fill the gap between these two extremes by analyzing a large variety of graphs with intermediate connectivity, thus providing insight into the relationship between the connectivity structure of the network and the version age, and uncovering a network connectivity--information freshness tradeoff.
\end{abstract}

\section{Introduction}\label{sec1}
With the roll-out of 5G communication, and a focus on decentralized wireless communication systems in 6G, including seamless machine-machine and human-machine interactions for time-critical tasks, timeliness has become an important metric to consider. This is of importance in many upcoming applications, such as drone swarms, networks of self-driving cars, smart manufacturing systems, and remote medical procedures. 

It is now known that latency alone is not sufficient to characterize timeliness \cite{popovski2022perspective}, and new metrics are needed. One such metric is age of information \cite{kaul2012real,sun2019age, yatesJSACsurvey}. Age of information is a well-accepted metric when studying the freshness of information in a wireless network\cite{banerjee2023re}. Several related freshness metrics are age of incorrect information \cite{maatouk20AOII}, age of synchronization \cite{zhong18AoSync}, binary freshness metric \cite{cho3BinaryFreshness}, and version age of information \cite{yates21gossip, Abolhassani21version, melih2020infocom}.

\begin{figure}[t]
    \centering
    \includegraphics[width = 0.7\linewidth]{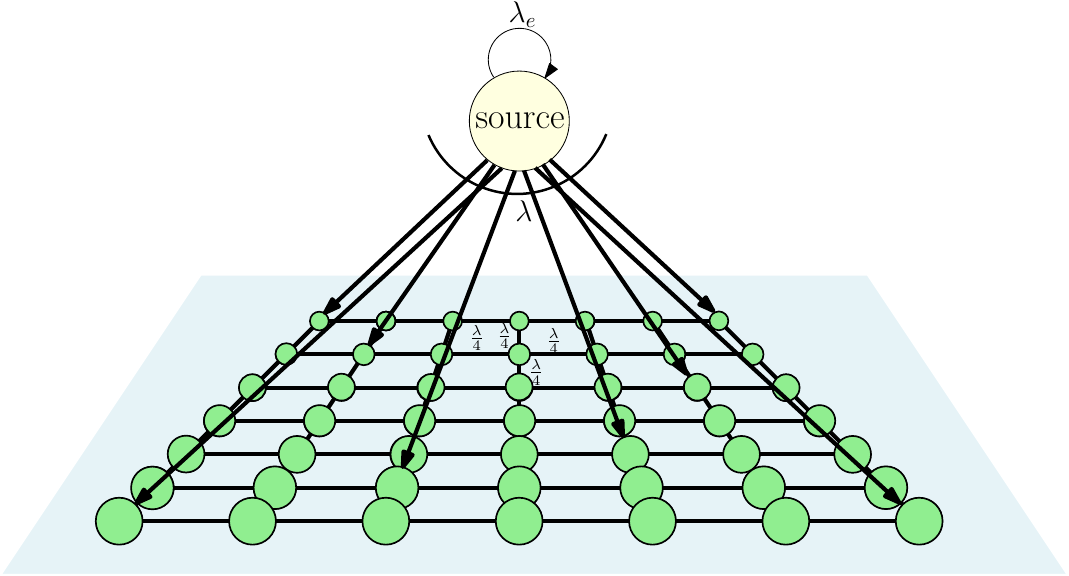}
    \caption{A gossip network where the node in light yellow is the source generating updates and sending them to a network of nodes connected in a two-dimensional grid. There are no boundaries in the grid, and all nodes have four neighbors.}
    \label{fig_grid_network}
\end{figure}

Gossip has been studied extensively in computer science. The first paper to study gossip algorithms is \cite{demers87epidemic_gossip}. Since then, many works\cite{yates21spawc, yaron03thesis, vocking2000, pittel_87_gossip, deb2006AlgebraicGossip, devavrat2006, Sanghavi2007GossipFileSplit}, have studied gossip networks from the context of disseminating information or performing a network-wide operation as fast as possible. In this paper, we consider gossip in networks as a means to disseminate information between nodes in the network, so that each node follows the update at a source node as closely as possible. To analyze such a network, we consider the version age of information metric. In simple words, the version age of information of a node in a network is the number of versions behind the node is with respect to the source node which has the latest version of the update. Gossip was used with version age as a metric for the first time in \cite{yates21gossip}. \cite{yates21gossip} characterizes the version age of information for general networks with Poisson update times and describes a set of recursive equations that can be used to calculate the average version age of connected subsets of a network. \cite{yates21gossip} also shows that the average version age of a single node in a fully-connected network scales as $O(\log{n})$ with the network size $n$. \cite{yates21gossip} experimentally observes, and \cite{buyukates21CommunityStructure} mathematically shows, that for a ring network, the version age scales as $O(\sqrt{n})$. \cite{kaswan22timestomp} studies a network with a timestomping adversary, which can change the timestamps of the updates and fool the nodes into accepting an older version of the update. \cite{kaswan22jamming} studies the metric in the case where there are jamming adversaries. \cite{kaswan23nonpoissonISIT, kaswan23nonpoissonCDC} study the version age of information in a non-Poisson update setting. \cite{mitra_allerton22} considers version age in an age-sensing multiple access channel. \cite{mitra2023timely} considers opportunistic gossiping protocols that achieve $O(1)$ scaling for version age in distributed multiple access channels. \cite{abd2023distribution} studies the distributions of version age and its moments. \cite{pappas2023versionenergyharvesting} studies a gossip network consisting of energy harvesting sensors with the version age metric. \cite{bastopcu2022role} considers a gossip network where nodes decide the new update received using majority rule, and introduces a new error metric similar to the binary freshness metric.

In this paper, we extend the analysis of version age in gossip networks to include several structured graph models. In order to do so, we modify the recursive equations in \cite{yates21gossip} and provide an upper bound and a lower bound for the average version age of a set in terms of the neighbors of a set or the neighboring edges of a set. Using these upper bound recursive equations, and lower bounds on the number of incoming edges to a set, we find an upper bound for the average version age of a single node for various graphs.

\begin{figure}[t]
    \centering
    \includegraphics[width = 0.7\linewidth]{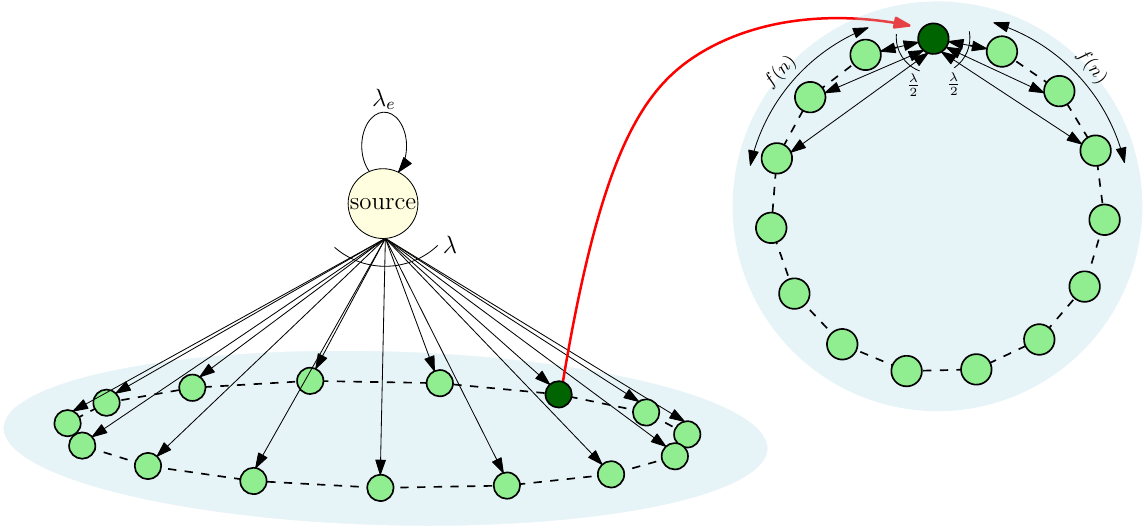}
    \caption{A gossip network where the gossiping nodes form a generalized ring network. The source node updates itself at rate $\lambda_e$, and disseminates information to nodes arranged in a ring at rate $\lambda$. Each node communicates with $f(n)$ nodes on each of its side, thus communicating with $2f(n)$ nodes in total.}
    \label{fig_ring_network}
\end{figure}

\begin{figure}[t]
    \centering
    \includegraphics[width = 0.9\linewidth]{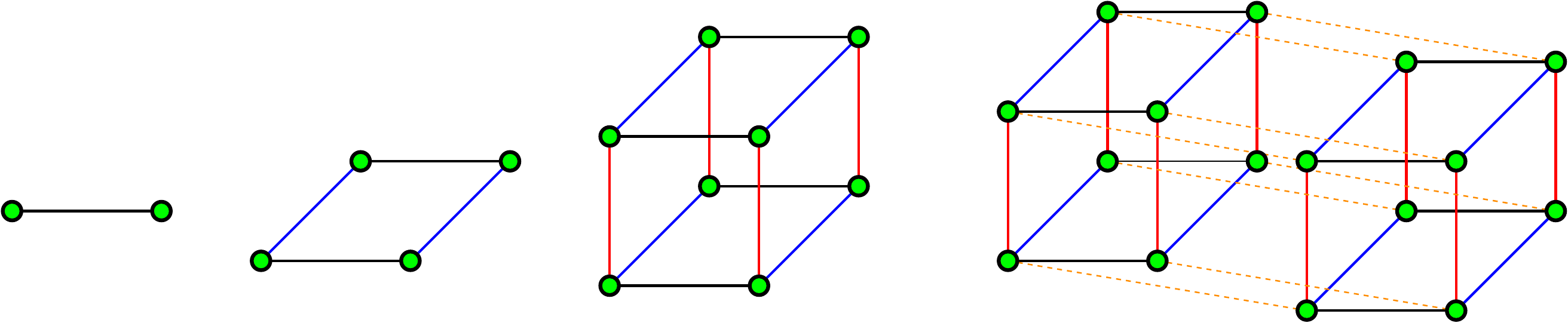}
    \caption{The evolution of a unit hypercube network, from one dimension to four dimensions.}
    \label{fig_unit_hypercube_network}
\end{figure}

\begin{figure}[t]
    \centering
    \includegraphics[width = 0.4\linewidth]{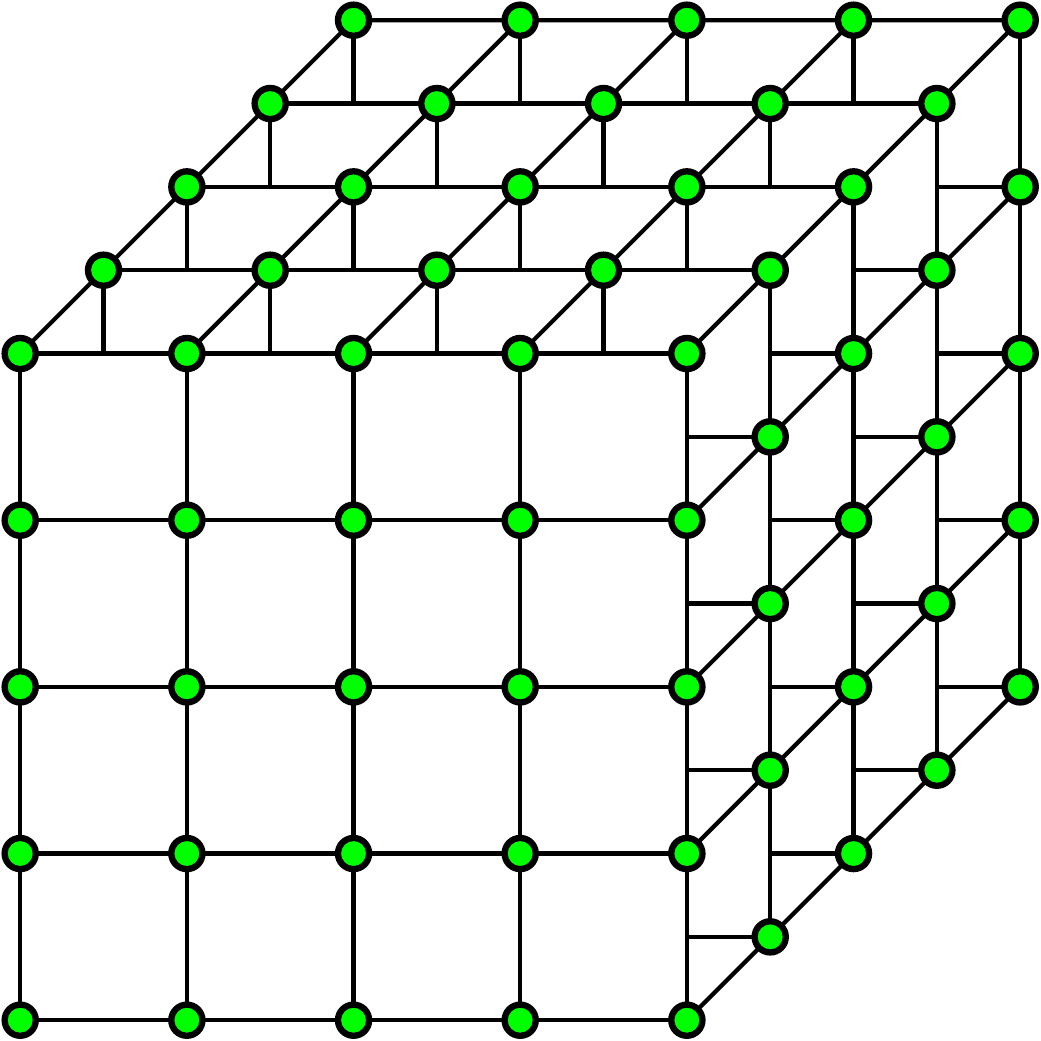}
    \caption{A three-dimensional hypercube network with side length five.}
    \label{fig_general_hypercube_network}
\end{figure}

In this work, we consider the following graph structures, and obtain the following results:
\begin{enumerate}
    \item A two-dimensional grid graph of $n$ nodes with general sides $k$ and $m$ such that $n=km$ (see Fig.~\ref{fig_grid_network}): We find that if $k = \Omega(\sqrt{m})$, then the version age scales as $O((km)^{\frac{1}{3}})$, i.e., as $O(n^{\frac{1}{3}})$. On the other hand, if $k = o(\sqrt{m})$, then the version age scales as $O(km^\frac{1}{2})$, i.e., as $O(n^\frac{1}{2})$.
    \item A generalized ring network, where nodes are placed in a ring structure, and each node is connected to $f(n)$ nodes on either side (see Fig.~\ref{fig_ring_network}): We find that the version age scales as $O\left(\log{f(n)} + \sqrt{\frac{n}{f(n)}}\right)$.
    \item A unit hypercube network in dimension $\log{n}$ (see Fig.~\ref{fig_unit_hypercube_network}): We find that the version age scales as $O(\log{n}\log{\log{n}})$.
    \item A general hypercube in fixed $d$-dimensional space, where each side goes to infinity (see Fig.~\ref{fig_general_hypercube_network}): We perform simulations and see that the version age scales as $O(n^{\frac{1}{d+1}})$. We provide conjectures to support this claim.
\end{enumerate}

\section{System Model and the Version Age Metric}\label{sec2}
We consider a wireless network which has one source node generating or observing updates of a particular process. This source updates itself as a rate $\lambda_e$ Poisson process independent of all other processes in the network. The network also contains a gossip network, denoted by $\mathcal{N}$, that consists of $n$ nodes. The source node shares the updates with the gossip network as a Poisson process with a total rate $\lambda$, giving each node in the network an equal chance of being updated. This can be understood alternatively as, the source sends updates to each node as a Poisson process of rate $\frac{\lambda}{n}$, independent of all other processes.

To quantify the freshness of update versions at the nodes in the gossip network, we use the version age metric. We start by defining the counting process associated with the updates at the source as $N_0(t)$. For node $i$ in the gossip network, we define the counting process associated with the updates as $N_i(t)$. Next, we define the version age of node $i$ as $X_i(t) = N_0(t) - N_i(t)$. Hence, the version age of any node in the gossip network is the number of versions the particular node is behind with respect to the source node. Similarly, we define the version age of a connected subset $S$ of the gossip network as $X_S(t) = \min_{j \in S} X_j(t)$. Note that $X_S(t)$ is the smallest version age in the subset of nodes in $S$. Finally, we define the limiting average version age of set $S$ as $v_S = \lim_{t \rightarrow \infty} \mathbb{E}[X_S(t)]$. 

The version age of a node evolves as follows: If the source node updates itself, then the version age of every node in the gossip network increases by $1$. If the source node sends an update to a node, then the node's version age reduces to $0$. If a node $i$ sends its version of the update to node $j$, then $j$ updates itself if it has an older version than the one sent by $i$. Otherwise, it keeps its own version and rejects $i$'s version. 

\begin{figure}[t]
    \centering
    \includegraphics[width = 0.55\linewidth]{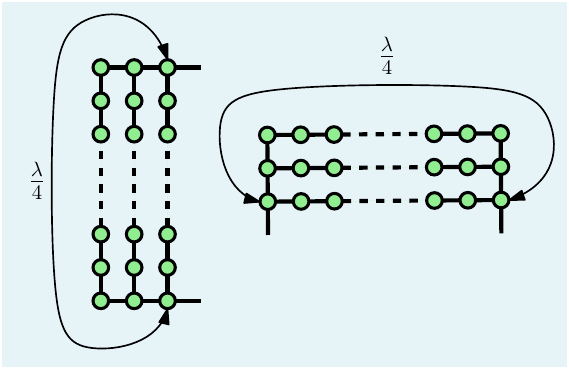}
    \caption{All nodes have four connections. Nodes that are seemingly at the boundary in Fig.~\ref{fig_grid_network} are connected in a wrap-around fashion to the nodes on the opposite side of the row or column.}
    \label{fig_grid_cross_connections}
\end{figure}

We define the rate of information flow from node $i$ to node $j$ as $\lambda_{ij}$ which is the rate of the Poisson update process from $i$ to $j$. We say that node $i$ is a neighboring node of set $S$ if $\lambda_{ij} > 0$ for some $j \in S$. We define the total rate of information flow into the set $S$ from neighboring node $i$ as $\lambda_i(S)$, 
\begin{align}
    \lambda_i(S) = \begin{cases}
        \sum_{j \in S} \lambda_{ij}, & i \notin S\\
        0, & \text{otherwise}.
    \end{cases}
\end{align}
Similarly, $\lambda_0(S)$ is the total rate of information flow from the source node into the set $S$. We define $N(S)$ to be the set of nodes which are neighboring nodes to the set $S$. Finally, we define an incoming edge to set $S$ as an edge that emanates from a neighboring node of set $S$ and ends at a node in $S$, and the set of all incoming edges to $S$ is $E(S)$.

With these definitions, we are now equipped to use the recursive equations for general networks proposed in \cite{yates21gossip}. We will also use these in Section~\ref{sec3} to find upper and lower bounds for $v_S$ by modifying the recursive equations. In later sections, we use these to find upper bounds for the version age for the following different network topologies.

\begin{figure}[t]
    \centering
    \includegraphics[width = 0.45\linewidth]{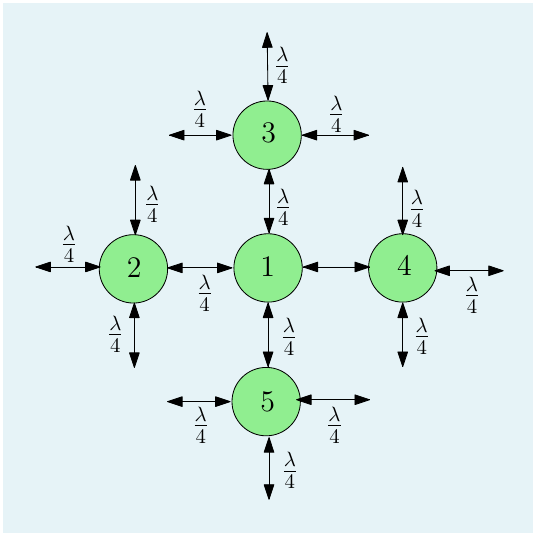}
    \caption{Each node updates its four neighbors with an update rate of $\frac{\lambda}{4}$. Total update rate per node is $\lambda$.}
    \label{fig_grid_rates}
\end{figure}

\paragraph{Description of the Two-Dimensional Grid Network:}
The network of gossiping nodes is arranged in a two-dimensional rectangular grid structure, with there being $m$ nodes in the longer side, and $k$ nodes in the shorter side. Each node in the grid has four neighbors, in the sense that there are no boundaries in the network, i.e., a  node that is seemingly at the boundary in Fig.~\ref{fig_grid_network} is connected in a wrap-around fashion to the nodes on the opposite side of the row or column as shown in Fig.~\ref{fig_grid_cross_connections}. Each node has a total gossiping rate of $\lambda$ which it divides into four equal parts to gossip with its neighbors. Hence, each node in the grid gossips with its four neighbors in a push-style gossiping protocol as a rate $\frac{\lambda}{4}$ Poisson process, as shown in Fig.~\ref{fig_grid_rates}, independent of all other processes in the network.

\paragraph{Description of the Generalized Ring Network:}
Each node in the generalized ring network gossips with its neighbors, which are the nearest $2f(n)$ nodes, i.e., it gossips with $f(n)$ nodes on each side. It sends its version of the update to each neighboring node as a rate $\frac{\lambda}{2f(n)}$ Poisson process, resulting in a total rate $\lambda$ for dissemination of information. This is shown in detail on the right hand side in Fig.~\ref{fig_ring_network}.

\paragraph{Description of the Unit Hypercube Network:}
The nodes in the unit hypercube network are arranged in the shape of an $m$-dimensional hypercube, thus having $n=2^m$ nodes in total. According to the properties of a hypercube, each node gossips with $m$ neighbors as a rate $\frac{\lambda}{m}$ Poisson process. That is, each of the $n$ nodes in the network gossips with $m=\log n$ neighbors. We analyze this network as the dimension of the unit hypercube $m$ goes to infinity (hence, the number of nodes $n$ goes to infinity). In Fig.~\ref{fig_unit_hypercube_network}, we show how the sets evolve as $m$ increases from  $1$ to $4$, and $n=2^m$ correspondingly increases from $2$ to $16$.

\paragraph{Description of a General $d$-dimensional Hypercube Network:}
The nodes in a general hypercube network are arranged in the shape of a hypercube in $d$ dimensions, where $d > 2$. The hypercube has side length $m$, unlike the previous case, where each side was unit length. Hence, there are $n=m^d$ nodes in the network. We assume that this network has wrapped-around edges like the two-dimensional grid network. Hence, each node gossips with $2m$ neighbors as a rate $\frac{\lambda}{2m}$ Poisson process. We analyze this network as the length of the side $m$ goes to infinity, and the dimension of the hypercube $d$ is constant. In Fig.~\ref{fig_general_hypercube_network}, we show a three dimensional hypercube ($d=3$) with side length five ($m=5$) with $n=m^d=125$ nodes. Note that a $d=2$ dimensional hypercube is a square two-dimensional grid, and a $d=1$ dimensional hypercube is a line. A line is a ring that is cut in the middle.

\section{The Recursive Equations and Upper/Lower Bounds}\label{sec3}
In \cite{yates21gossip}, Yates developed a recursive formula to find the version age of a connected set $S$ in a general network as,
\begin{align}
    v_S = \frac{\lambda_e + \sum_{i \in N(S)}\lambda_i(S)v_{S \cup \{i\}}}{\lambda_0(S) + \sum_{i \in N(S)}\lambda_i(S)}. \label{yates-recursion}
\end{align}
This recursive formula expresses the version age of a subset $S$ as a linear combination of the version ages of subsets that are obtained by adding one more neighboring node to this subset, i.e., $S \cup \{i\}$, for all possible $i$. For a general network, the number of such equations obtained is exponential in the number of nodes, and is difficult to solve in closed-form.

We want to find the version age of a typical node $j$, $v_{\{j\}}$, in certain structured networks such as grids, generalized rings, and hypercubes. From the symmetry of the wrapped-around networks we consider, all nodes in the two-dimensional grid network and the $d$-dimensional general hypercube will have the same version age. Each node in the generalized ring network and the unit hypercube network will also have the same version age, since the networks are symmetric with respect to each node. Towards that end, we need to start with a single node $j$, i.e., $S=\{j\}$ in \eqref{yates-recursion}, and keep adding neighboring nodes $i$ until the entire network is exhausted. We know that the version age of the entire network $v_{\mathcal{N}}$ is $v_{\mathcal{N}}=\frac{\lambda_e}{\lambda}$. In principle, working backwards, we can find the version age of node $j$, $v_{\{j\}}$, by solving all of the linear equations obtained through the recursion in \eqref{yates-recursion}. To achieve that, we need to identify and exploit simplifications/symmetries in these equations.

References \cite{yates21gossip} and \cite{buyukates21CommunityStructure} were able to find the version ages of fully-connected and ring networks, respectively, by developing relatively simple bounds, as in those cases, the age of a subset depends only on the size of the subset due to the symmetry of the networks. However, in the case of a more complex network, subsets of the same size may have widely varying ages. Consider for instance a subset of $9$ nodes in a two-dimensional grid network: those $9$ nodes could be on a line, versus on a $3\times 3$ square, versus on an $L$ shape with $6$ nodes on the longer arm of $L$ and $4$ nodes on the shorter arm. All these subsets of $9$ nodes, even in a symmetric wrapped-around grid network, will have different version ages.

To tackle this problem, we find a general upper bound (Lemma~\ref{lemma1}) and a general lower bound (Lemma~\ref{lemma2}) for $v_S$ in a general network. In later sections, we show how to apply these lemmas to find upper bounds for the complex networks discussed in Section~\ref{sec2}.

The recursion in \eqref{yates-recursion} provides a method to find the version age of any connected subgraph of a gossip network. Even for the simplest graphs, such as rings and fully-connected networks, the application of this recursion is tedious. For more complex graphs, such as the ones we consider in this paper, use of symmetry is restricted and direct application of the recursion is challenging since the number of sets grows quickly. In order to simplify the application of the recursion, the following lemma modifies the recursion to find bounds for sets only based on the one-expanded set that has the highest version age as opposed to all one-expanded sets. This enables us to use the geometry of the network and restrict the number of sets we need to handle to find a bound for the set in question.

\begin{lemma} \label{lemma1}
    In any general gossip network, the following upper bound holds for a subset of nodes $S$,
    \begin{align}
        v_S \leq \frac{\lambda_e + |N(S)| \cdot \min_{i\in N(S)}{\lambda_i(S)} \cdot \max_{i\in N(S)}{v_{S \cup \{i\}}}}{\lambda_0(S) + |N(S)| \cdot \min_{i\in N(S)}\lambda_i(S)}.  \label{eq_ub}
    \end{align}
\end{lemma}

\begin{Proof}
    We start by rearranging \eqref{yates-recursion} as follows,
    \begin{align}
        \lambda_e = \lambda_0(S)v_S + \sum_{i \in N(S)}\lambda_i(S)(v_S - v_{S \cup \{i\}}).  \label{rearranged_recursion}
    \end{align}
    Now, we lower bound the sum on the right hand side as,
    \begin{align}
        \lambda_e &\geq \lambda_0(S)v_S + |N(S)|\min_{i \in N(S)}\lambda_i(S)(v_S - v_{S \cup \{i\}})\\
        &\geq \lambda_0(S)v_S + |N(S)|\min_{i\in N(S)}\lambda_i(S) \min_{i \in N(S)}(v_S - v_{S \cup \{i\}})   \\
        &= \lambda_0(S)v_S + |N(S)|\min_{i \in N(S)}\lambda_i(S)(v_S  - \max_{i \in N(S)} v_{S \cup \{i\}}).   \label{eq7}
    \end{align}
    Rearranging \eqref{eq7} gives the desired result.
\end{Proof}

In a similar way, we can also prove the lower bound in the following lemma.

\begin{lemma} \label{lemma2}
    In any general gossip network, the following lower bound holds for a subset of nodes $S$,
    \begin{align}
        v_S \geq \frac{\lambda_e + |N(S)| \cdot \max_{i\in N(S)}{\lambda_i(S)} \cdot \min_{i\in N(S)}{v_{S \cup \{i\}}}}{\lambda_0(S) + |N(S)| \cdot \max_{i\in N(S)}\lambda_i(S)}. \label{eq_lb}
    \end{align}
\end{lemma}

\begin{Proof}
    We start with \eqref{rearranged_recursion}, and this time, we write an upper bound for the right hand side,
    \begin{align}
    \lambda_e &\leq \lambda_0(S)v_S + |N(S)|\max_{i \in N(S)}\lambda_i(S)(v_S - v_{S \cup \{i\}})\\
    &\leq \lambda_0(S)v_S + |N(S)|\max_{i\in N(S)}\lambda_i(S) \max_{i \in N(S)}(v_S - v_{S \cup \{i\}})   \\
    &= \lambda_0(S)v_S + |N(S)|\max_{i \in N(S)}\lambda_i(S)(v_S  - \min_{i \in N(S)} v_{S \cup \{i\}}).   \label{eq8}
    \end{align}
    Rearranging \eqref{eq8} gives the desired result.
\end{Proof}

\section{Version Age in an $m \times k$ Grid}\label{sec4}
In this section, we find an upper bound for the version age of a single node in the two-dimensional grid network, described in Section~\ref{sec2}. We note that the average version age of each node in the grid will be the same due to the symmetry of the network. In order to calculate the upper bound, we will modify the recursive equations of \cite{yates21gossip}.

In the following lemma, we find an upper bound for $v_S$ in terms of the number of incoming edges, instead of the number of neighbors of $S$. Then, we find a lower bound for the number of incoming edges for subsets of the grid network, and write an upper bound for $v_S$ in terms of the number of nodes in $S$.

\begin{lemma}\label{lemma3}
    In the grid network, consider a connected subgraph $S$. Suppose $|S| = j$ such that $j \leq \frac{k^2}{4}$. Then, we have,
    \begin{align}
        v_S \leq \frac{\frac{2\lambda_e}{\lambda} + \sqrt{j} \max_{i \in N(S)}v_{S \cup \{i\}}}{\frac{j}{mk} + \sqrt{j}}. \label{grid_recursion_eq1}
    \end{align}
    Further, if $\frac{k^2}{4}+1 \leq j \leq mk-\frac{k^2}{4}$, then, we have,
    \begin{align}
        v_S \leq \frac{\frac{2\lambda_e}{\lambda} + k\max_{i \in N(S)}v_{S \cup \{i\}}}{\frac{j}{mk} + k}.\label{grid_recursion_eq2}
    \end{align}
\end{lemma}

 \begin{figure}
        \centering
        \includegraphics[width = 0.27\linewidth]{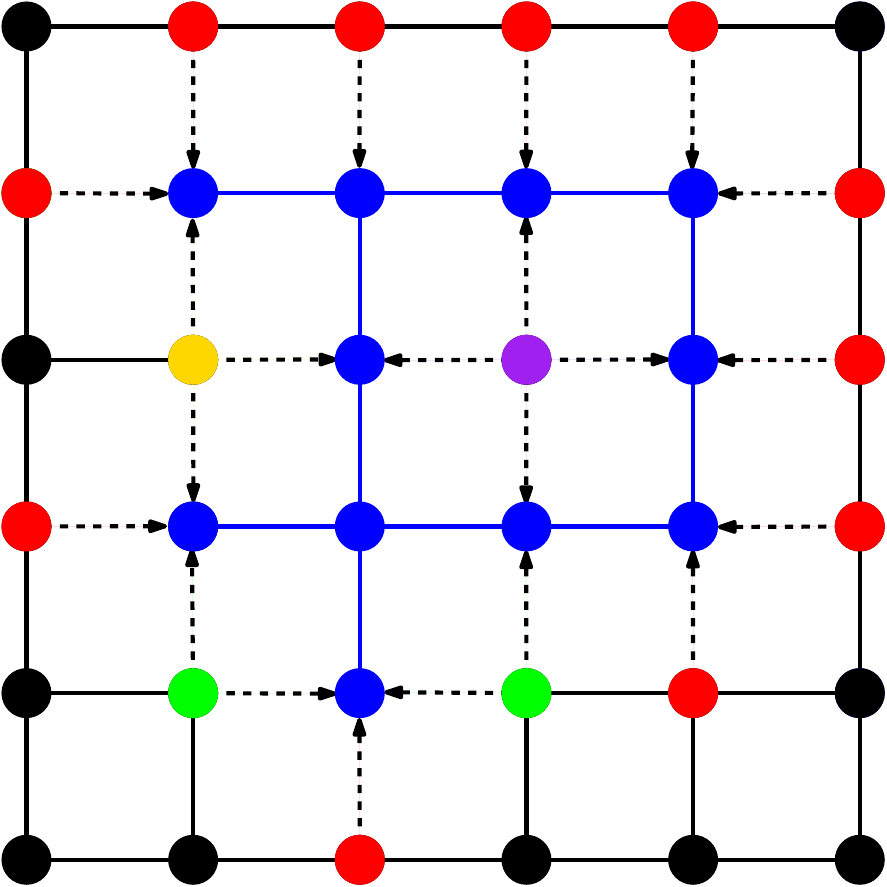}
        \caption{An illustrative example for sets in \eqref{eq_edge_types_grid_A}, \eqref{eq_edge_types_grid_B}, \eqref{eq_edge_types_grid_C} and \eqref{eq_edge_types_grid_D}. The set $S$ is colored blue. Nodes of type $A$ are colored red (they update into $S$ with one link), nodes of type $B$ are colored green (update into $S$ with two links), nodes of type $C$ colored yellow (update into $S$ with three links) and nodes of type $D$ colored purple (update into $S$ with four links).}
        \label{fig_edge_types_grid}
\end{figure}

\begin{Proof}
    We start with \eqref{rearranged_recursion},
    \begin{align}
       \lambda_e = \lambda_0(S)v_S + \sum_{i \in N(S)}\lambda_i(S)(v_S - v_{S \cup \{i\}}). \label{recur_eq}
    \end{align}
    Now, we define a function $E_S(i) = \sum_{a \in S} \mathbb{I}(\lambda_{ia} > 0)$, where $i \in N(S)$ and $\mathbb{I}(\cdot)$ is the indicator function. $E_S(i)$ is the number of incident edges on set $S$ that emanate from node $i$. Then, we can partition $N(S)$ as follows,
    \begin{align}
        A = \{i \in N(S): E_S(i) = 1\} \label{eq_edge_types_grid_A},\\
        B = \{i \in N(S): E_S(i) = 2\} \label{eq_edge_types_grid_B},\\
        C = \{i \in N(S): E_S(i) = 3\} \label{eq_edge_types_grid_C},\\
        D = \{i \in N(S): E_S(i) = 4\} \label{eq_edge_types_grid_D}.
    \end{align}   
    That is, $A$ is the set of neighbors of $S$ that update into the set $S$ with a single link (single edge), $B$ is the set of neighbors of $S$ that update into the set $S$ with two links (two edges), and similarly, $C$ with three links (three edges), and $D$ with four links (four edges). An illustrative example is shown in Fig.~\ref{fig_edge_types_grid}. Using these sets, we bound \eqref{recur_eq} as,
    \begin{align}
        \lambda_e =& \lambda_0(S)v_S + \sum_{i \in A}\frac{\lambda}{4}(v_S - v_{S \cup \{i\}}) + \sum_{i \in B}\frac{\lambda}{2}(v_S - v_{S \cup \{i\}})+ \sum_{i \in C}\frac{3\lambda}{4}(v_S - v_{S \cup \{i\}}) \notag\\
        &+ \sum_{i \in D}\lambda(v_S - v_{S \cup \{i\}})\\
        \geq &\lambda_0(S)v_S + |A|\frac{\lambda}{4}\min_{i \in A}(v_S - v_{S \cup \{i\}})+ |B|\frac{\lambda}{2}\min_{i \in B}(v_S - v_{S \cup \{i\}})+ |C|\frac{3\lambda}{4}\min_{i \in C}(v_S - v_{S \cup \{i\}}) \notag\\
        &+ |D|\lambda\min_{i \in D}(v_S - v_{S \cup \{i\}})\\
        \geq &\lambda_0(S)v_S + |A|\frac{\lambda}{4}\min_{i \in N(S)}(v_S - v_{S \cup \{i\}}) + |B|\frac{\lambda}{2}\min_{i \in N(S)}(v_S - v_{S \cup \{i\}})+ |C|\frac{3\lambda}{4}\min_{i \in N(S)}(v_S - v_{S \cup \{i\}}) \notag\\
        &+ |D|\lambda\min_{i \in N(S)}(v_S - v_{S \cup \{i\}})\\
        =& \lambda_0(S)v_S+ \frac{\lambda}{4}(|A| + 2|B| + 3|C| + 4|D|)\min_{i \in N(S)}(v_S  -  v_{S \cup \{i\}})\\
        =& \lambda_0(S)v_S+ \frac{\lambda}{4}(|A| + 2|B| + 3|C| + 4|D|)(v_S - \max_{i \in N(S)}v_{S \cup \{i\}}) . \label{eq_edgelb}
    \end{align}
    Now, we note that $|E(S)| = |A| + 2|B| + 3|C| + 4|D|$, where $|E(S)|$ is the total number of incoming edges into the set $S$. Thus, \eqref{eq_edgelb} states that for the two-dimensional grid, we have
    \begin{align}
     \lambda_e \geq \lambda_0(S)v_S + \frac{\lambda}{4} |E(S)| (v_S - \max_{i \in N(S)}v_{S \cup \{i\}}). \label{eq_edgelb_1}
    \end{align}

    Next, we use a result from \cite{harary1976extremal} and \cite{nowzari19improved} to provide a lower bound for the number of incoming edges into the set $S$, i.e., $|E(S)|$, to further lower bound \eqref{eq_edgelb_1}. According to \cite{harary1976extremal}, on an infinite two-dimensional grid, of all the connected subsets with a fixed size $j$, the subset that has the minimum number of incoming edges into the subset is the \emph{spiral}. The number of incoming edges into a spiral of $j$ nodes is given by $2\lceil 2\sqrt{j} \rceil$. This bound has also been used in \cite{nowzari19improved} in the analysis of gossip algorithms for max-consensus on grids. Hence, we have the following lower bound for $|E(S)|$: 
    \begin{align}
        |E(S)| \geq 2\lceil 2\sqrt{j} \rceil \geq 4\lfloor \sqrt{j} \rfloor \geq 2 \sqrt{j}. \label{ES-bound}
    \end{align}
    An illustrative example for this bound is shown in Fig.~\ref{fig_different_edge_sets_grid}.

    \begin{figure}
        \centering
        \includegraphics[width = \linewidth]{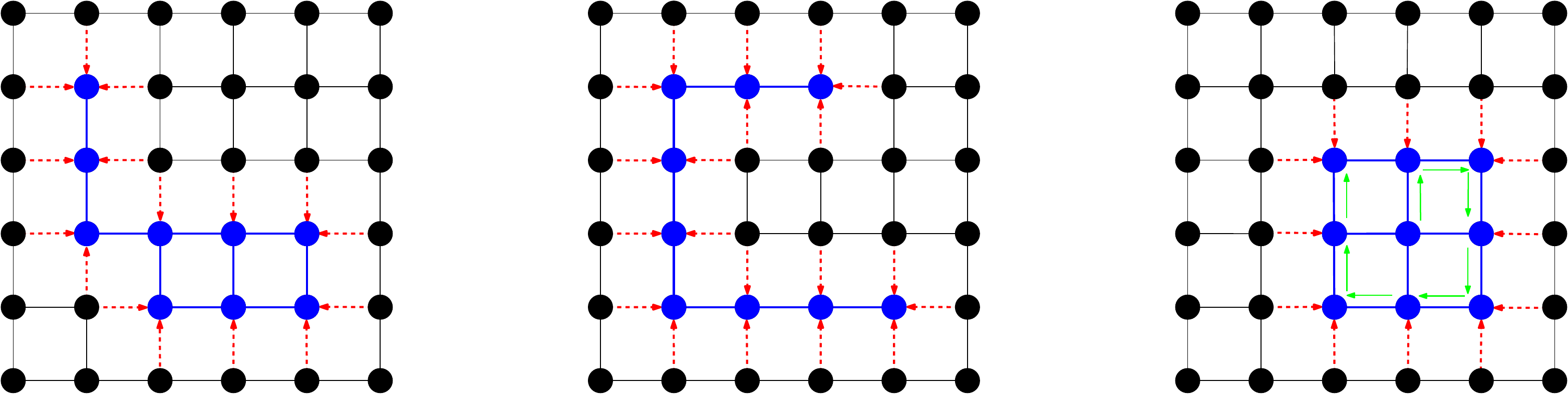}
        \caption{An illustration of the lower bound: $|E(S)| \geq 2\lceil 2\sqrt{j} \rceil$. The number of nodes in $S$ (blue colored) is $j=9$, thus the lower bound is $|E(S)|\geq 12$. Different types of sets with $9$ nodes have widely different number of incoming edges: The set on the left has $16$ incoming edges (red arrows), the set in the middle has $20$ incoming edges, and the set on the right has $12$ incoming edges. The set on the right is a \emph{spiral}, which is the set with the least number of incoming edges; the green lines show how the spiral is formed starting from the center node.}
        \label{fig_different_edge_sets_grid}
    \end{figure}

    However, our grid network is a finite network wrapped-around at the boundaries. Hence, the effects of the boundaries affect this bound once $j$ becomes large enough in comparison to the network size $n$. To see this, consider the sets on the grid given in Fig.~\ref{fig_grid_special_sets}. Both of the sets have $j=15$ nodes. Thus, the infinite-network lower bound given for these sets is $|E(S)| \geq 2\lceil 2\sqrt{j} \rceil=16$. We note that the right set has $16$ incoming edges. The right set, in fact, is a spiral. Note that the right set does not touch (i.e., make use of) the finite boundaries of the grid network. The left set, on the other hand, makes use of the boundaries in order to cut out the incoming edges on the extreme right and extreme left of the set. Hence, the left set has $13$ incoming edges, which is smaller than the infinite-network lower bound of $16$. Had the grid been infinite, the left set would have had $17$ incoming edges, one more than the spiral, satisfying the result in \cite{harary1976extremal}. We also observe that the set on the left has a lower bound on the number of incoming edges given by $2k$. Hence, we conclude that up to a certain size of the set, the spiral will have a lower number of incoming edges, and for sets larger than that we will have a smaller lower bound as in the set on the left in Fig.~\ref{fig_grid_special_sets}. The exact inflection point is given by comparing the two lower bounds,
    \begin{align}
        2k \geq 2\lceil 2\sqrt{j} \rceil.
    \end{align}
    This is satisfied for $j \leq \frac{k^2}{4}$. Hence, we conclude that $|E(S)|$ is bounded below by $2\lceil 2\sqrt{j} \rceil$, and therefore, by $2 \sqrt{j}$ from \eqref{ES-bound} by getting rid of the ceiling and floor functions, for $j$ up to $j = \frac{k^2}{4}-1$ and by $2k$ for $j$ between $\frac{k^2}{4}$ and $mk-\frac{k^2}{4}$. This is because we have assumed that $m\geq k$, hence the set will first be a spiral of size up to $\frac{k^2}{4}$. 

    \begin{figure}[t]
        \centering
        \includegraphics[width=0.7\linewidth]{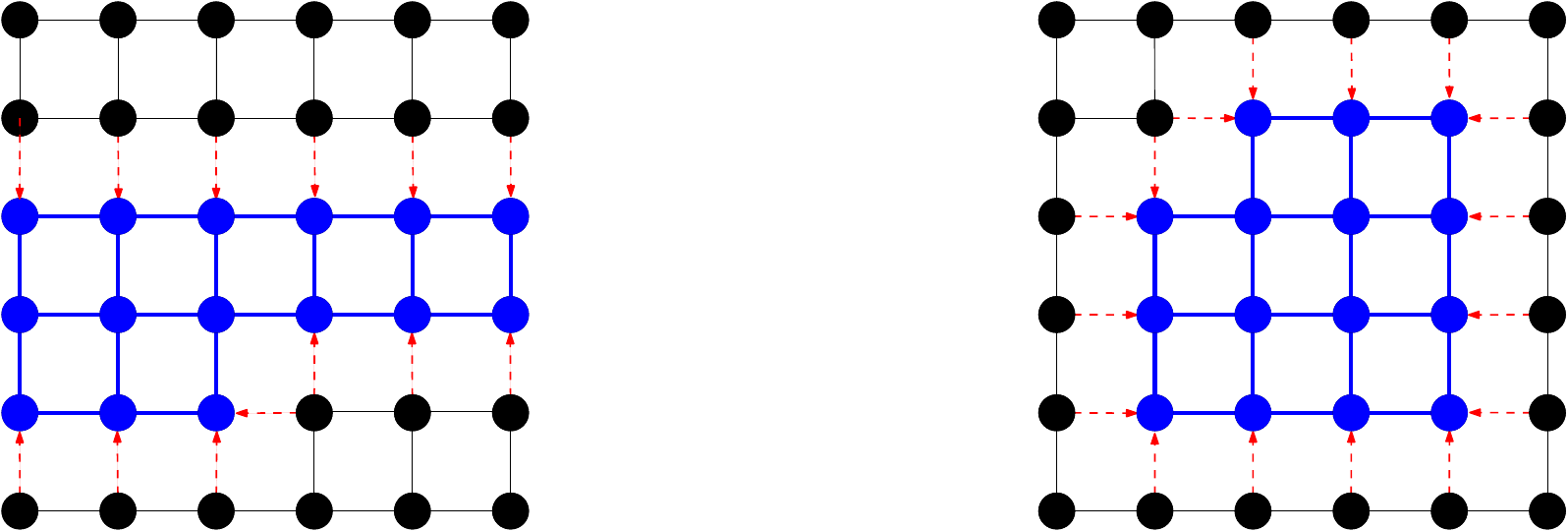}
        \caption{Comparison of the number of incoming edges of two sets with $15$ nodes on a $6\times 6$ grid network. Due to boundary constraints, the set on the left has less incoming edges than the spiral set on the right.}
        \label{fig_grid_special_sets}
    \end{figure}

    Thus, for $j \leq \frac{k^2}{4}$, substituting the first lower bound into \eqref{eq_edgelb_1}, we obtain
    \begin{align}
     \lambda_e \geq \lambda_0(S)v_S + \frac{\lambda}{4} 2 \sqrt{j} (v_S - \max_{i \in N(S)}v_{S \cup \{i\}}). \label{eq_edgelb_2}
    \end{align}
    Inserting $\lambda_0(S)=\frac{\lambda j}{mk}$ and rearranging \eqref{eq_edgelb_2}, we have
    \begin{align}
        v_S &\leq \frac{\lambda_e + \lambda \frac{\sqrt{j}}{2} \max_{i \in N(S)} v_{S \cup \{i\}}}{\frac{\lambda j}{mk} + \lambda \frac{\sqrt{j}}{2}}\\
        &= \frac{\frac{\lambda_e}{\lambda} + \frac{\sqrt{j}}{2}\max_{i \in N(S)} v_{S \cup \{i\}}}{\frac{j}{mk} + \frac{ \sqrt{j}}{2}}\\
        &= \frac{\frac{2\lambda_e}{\lambda} + \sqrt{j}\max_{i \in N(S)} v_{S \cup \{i\}}}{\frac{2j}{mk} + \sqrt{j}}\\
        &\leq \frac{\frac{2\lambda_e}{\lambda} + \sqrt{j}\max_{i \in N(S)} v_{S \cup \{i\}}}{\frac{j}{mk} + \sqrt{j}}.
    \end{align}
    
    Now, similarly, for the range $\frac{k^2}{4}+1 \leq j \leq mk-\frac{k^2}{4}$, we obtain
    \begin{align}
        v_S &\leq \frac{\lambda_e + \lambda \frac{k}{2} \max_{i \in N(S)} v_{S \cup \{i\}}}{\frac{\lambda j}{mk} + \frac{\lambda}{2} k}\\
        &= \frac{\frac{\lambda_e}{\lambda} + \frac{k}{2} \max_{i \in N(S)} v_{S \cup \{i\}}}{\frac{j}{mk} + \frac{k}{2}}\\
        &= \frac{\frac{2\lambda_e}{\lambda} +  k \max_{i \in N(S)} v_{S \cup \{i\}}}{\frac{2 j}{mk} + k}\\
        &\leq \frac{\frac{2\lambda_e}{\lambda} + k \max_{i \in N(S)} v_{S \cup \{i\}}}{\frac{j}{mk} + k},
    \end{align}
    which completes the proof.
\end{Proof}

This bound does not work for $j > mk-\frac{k^2}{4}$, since when $j$ increases beyond this, we have a new type of set with the least number of incoming edges on set $S$. This set $S$ is equivalent to the whole grid minus a spiral shaped portion which has nodes that are not in $S$. The spiral has the minimum number of incoming edges, and hence the same can be said about the complement set of nodes. This results in set $S$ having the least number of incoming edges. The bound for the case $j > mk-\frac{k^2}{4}$ can be concluded from,
\begin{align} \label{eq_higher_number_of_nodes}
    2\lceil 2\sqrt{mk-j}\rceil < 2k,
\end{align}
where $|S| = j$. We can now bound the age of sets with sizes between $mk-\frac{k^2}{4}$ and $mk$ and get
\begin{align}
    v_S \leq \frac{\frac{\lambda_e}{\lambda} + \lfloor \sqrt{mk-j} \rfloor \max_{i \in N(S)}v_{S \cup \{i\}}}{\frac{j}{mk} + \lfloor \sqrt{mk-j} \rfloor}
\end{align}
by a simple modification in Lemma~\ref{lemma3}.

We state and prove the first main result of this paper in Theorem~\ref{thm1} below.

\begin{theorem}\label{thm1}
    The version age of a single user in the grid network scales as $v_1=O(\sqrt{m})$ if $k = o(\sqrt{m})$, and otherwise, the version age scales as $v_1=O((mk)^{\frac{1}{3}})$.
\end{theorem}

\begin{Proof}
    In order to find the version age of a single user, we use \eqref{grid_recursion_eq1} and \eqref{grid_recursion_eq2} in Lemma~\ref{lemma3} recursively, and obtain an analytic expression.  The core idea we use to get recursive upper bounds using Lemma~\ref{lemma3} is as follows: Start with a single node as the set $S$. We can start with any node in the network, since each node will have the same average version age due to the symmetry of the network. Let us call the age of this set $v_1$. Then, using Lemma~\ref{lemma3}, we get an upper bound on $v_1$ in terms of the age of a set with the maximum version age which can be constructed by adding one neighboring node to $S$. The set which satisfies these conditions need not be known. We call the age of this set $v_2$. In an inductive fashion, suppose we have $|S|=j$ with age $v_j$, then we use Lemma~\ref{lemma3} to get an upper bound by constructing a set which includes $S$ and a neighboring node of $S$, such that the one-larger set has the highest version age of all the possible one-larger sets constructed by adding neighboring nodes. We call the age of this set $v_{j+1}$. Since the upper bound is formed by independently bounding $|E(S)|$ and $v_{S \cup \{i\}}$, and the bound on $|E(S)|$ is universal given the number of nodes in the set, we get the required recursion. Following this way of constructing one-larger sets, we eventually cover the entire grid, for which we know that the version age is exactly $\frac{\lambda_e}{\lambda}$. Hence, we recursively use these upper bounds to find an upper bound for the version age of a single node in the grid network. Thus, in essence, we analyze a recursion of the following form, to find a bound on $v_1$, by starting with $v_n=\frac{\lambda_e}{\lambda}$,    
    \begin{align}
        v_j \leq 
        \begin{cases}
        \dfrac{\frac{2\lambda_e}{\lambda} + \sqrt{j} v_{j+1}}{\frac{j}{mk} + \sqrt{j}}, & \quad j \leq \frac{k^2}{4},\\
        ~ & ~ \\
        \dfrac{\frac{2\lambda_e}{\lambda} + k v_{j+1}}{\frac{j}{mk} + k}, & \quad \frac{k^2}{4}+1 \leq j \leq mk-\frac{k^2}{4},\\
        ~ & ~ \\
        \dfrac{\frac{\lambda_e}{\lambda} + \lfloor \sqrt{mk-j} \rfloor v_{j+1}}{\frac{j}{mk} + \lfloor \sqrt{mk-j} \rfloor}, & \quad mk-\frac{k^2}{4} < j \leq mk.
        \end{cases} 
        \label{grid_recursion_eq}
    \end{align}
    
    Now, we write the complete recursion in a way similar to the one found in  \cite[(13)-(16)]{buyukates22ClusterGossip}. This sum contains $n+1$ terms, which we divide into three parts, corresponding to the terms numbered between $\{1, \ldots, \frac{k^2}{4}\}$, $\{\frac{k^2}{4}+1, \ldots, mk-\frac{k^2}{4}\}$ and $\{mk-\frac{k^2}{4}+1, \ldots, mk+1\}$. We call the sum of terms in each of these parts $X$, $Y$ and $Z$, respectively. First, we write a bound for the sum of terms making up $X$,
    \begin{align}
        X &\leq \frac{2\lambda_e}{\lambda}\left(\frac{1}{1+\frac{1}{mk}}\right)\left(1 + \sum_{i=1}^{\frac{k^2}{4}} \prod_{j=1}^i \frac{\sqrt{j}}{\sqrt{j+1} + \frac{j}{mk}}\right)\label{eq_floor_after}\\ 
        &\leq \frac{2\lambda_e}{\lambda}\left(1 + \sum_{i=1}^{\frac{k^2}{4}} \prod_{j=1}^i \frac{\sqrt{j}}{\sqrt{j+1} + \frac{j}{mk}}\right). \label{eq_approx}
    \end{align}
    Next, to find the order of the right hand side of \eqref{eq_approx}, we proceed as follows,
    \begin{align}
        X &\leq \frac{2\lambda_e}{\lambda}\left(1 + \sum_{i=1}^{\frac{k^2}{4}} \prod_{j=1}^i \frac{1}{(1+\frac{1}{j})^{\frac{1}{2}} + \frac{\sqrt{j}}{mk}}\right) \label{eq_binom_approx_before}\\ 
        &\leq \frac{2\lambda_e}{\lambda}\left(1 + \sum_{i=1}^{\frac{k^2}{4}} \prod_{j=1}^i \frac{1}{1 + \frac{1}{2}\frac{1}{j} - \frac{1}{8}\frac{1}{j^2} + \frac{\sqrt{j}}{mk}}\right), \label{eq_binom_approx_after}
    \end{align}
    where we used $(1+x)^{\frac{1}{2}}\approx 1 + \frac{1}{2} x - \frac{1}{8}x^2$, for small $x$. Now, we define,
    \begin{align}
        a_i &= \prod_{j=1}^i \frac{1}{1 + \frac{1}{2}\frac{1}{j} - \frac{1}{8}\frac{1}{j^2} + \frac{\sqrt{j}}{mk}}. \label{def_ak}
    \end{align}
    By taking $\log$ of both sides, and using $\log(1+x) \approx x$, for small $x$, we obtain,
    \begin{align}
        -\log{a_i} &=\sum_{j=1}^i \log\left(1+\frac{1}{2}\frac{1}{j} - \frac{1}{8}\frac{1}{j^2} + \frac{\sqrt{j}}{mk}\right)\\ 
        &\approx \sum_{j=1}^i \left(\frac{1}{2}\frac{1}{j} - \frac{1}{8}\frac{1}{j^2} + \frac{\sqrt{j}}{mk}\right)\\
        &\approx \frac{1}{2}(\log{i} + \gamma)- \delta + \frac{2i^{\frac{3}{2}}}{3mk}, \label{final_approx}
    \end{align}
    where $\gamma \approx 0.577$ is the Euler-Mascheroni constant, $\delta$ is a finite positive constant upper bounded by $\frac{\pi^2}{48}$ since $\sum_j \frac{1}{j^2} \rightarrow \frac{\pi^2}{6}$, and the third term is bounded by the first term of the Euler-MacLaurin series expansion. Substituting \eqref{final_approx} in \eqref{eq_approx},
    \begin{align}
        X &\leq \frac{2\lambda_e}{\lambda}\left(1 + \alpha\sum_{i=1}^{\frac{k^2}{4}} e^{-\frac{1}{2}\log{i} - \frac{2i^{\frac{3}{2}}}{3mk}}\right)\\
        &= \frac{2\lambda_e}{\lambda}\left(1 + \alpha\sum_{i=1}^{\frac{k^2}{4}} \frac{1}{i^{\frac{1}{2}}} e^{-\frac{2i^{\frac{3}{2}}}{3mk}}\right) \\
        &\leq \frac{2\lambda_e}{\lambda}\left(1 + \alpha\sum_{i=1}^{mk} \frac{1}{i^{\frac{1}{2}}} e^{-\frac{2i^{\frac{3}{2}}}{3mk}}\right)\label{eq_reimann_sum},
    \end{align}
    where $\alpha$ considers the constants multiplied to the summation. 
    
    Next, we define a function $h$, 
    \begin{align}
        h(x) = \frac{1}{x^{\frac{1}{2}}}e^{-\frac{2}{3}x^{\frac{3}{2}}},
    \end{align}
    and then, write the Riemann sum with step size $(mk)^{-\frac{2}{3}}$,
    \begin{align}
        \sum_{i=1}^{mk} \frac{1}{(mk)^{\frac{2}{3}}} h\left(\frac{i}{(mk)^{\frac{2}{3}}}\right) = \int_{0}^{\infty} \frac{1}{t^{\frac{1}{2}}} e^{-\frac{2}{3}t^{\frac{3}{2}}}dt = \beta < \infty. \label{beta}
    \end{align}
    Hence, we continue as follows,
    \begin{align}
        \sum_{i=1}^{mk} \frac{1}{(mk)^{\frac{2}{3}}} h\left(\frac{i}{(mk)^{\frac{2}{3}}}\right) = \frac{1}{(mk)^{\frac{1}{3}}} \sum_{i=1}^{mk} \frac{1}{i^{\frac{1}{2}}} e^{-\frac{2i^{\frac{3}{2}}}{3mk}} = \beta.
    \end{align}
    Finally, continuing from \eqref{eq_reimann_sum}, we have,
    \begin{align}
        X \leq \frac{2\lambda_e}{\lambda}\left(1 + \beta'(mk)^{\frac{1}{3}}\right), \label{final_eq}
    \end{align}
    concluding that $X = O((mk)^{\frac{1}{3}})$.

    Next, we write the bound for the sum of terms making up $Y$,
    \begin{align}
        Y &\leq \frac{2\lambda_e}{\lambda}\left( \prod_{j=1}^{\frac{k^2}{4}}\frac{\sqrt{j}}{\frac{j}{mk} + \sqrt{j+1}}\right)\left( \sum_{i=\frac{k^2}{4}+1}^{mk-\frac{k^2}{4}}\prod_{j=\frac{k^2}{4}+1}^i \frac{1}{1+\frac{j}{mk^2}}\right)\\
        &= \frac{2\lambda_e}{\lambda}\left( \prod_{j=1}^{\frac{k^2}{4}}\frac{\sqrt{j}}{\frac{j}{mk} +  \sqrt{j+1} }\right)\left( \prod_{j=1}^{\frac{k^2}{4}} \frac{1}{1+\frac{j}{mk^2}}\right)^{-1}\left( \sum_{i=\frac{k^2}{4}+1}^{mk-\frac{k^2}{4}}\prod_{j=1}^i \frac{1}{1+\frac{j}{mk^2}}\right).
    \end{align}
    The first product here is just $a_{\frac{k^2}{4}}$ from \eqref{def_ak}, and we use this substitution,
    \begin{align}
        Y &\leq \frac{2\lambda_e}{\lambda}\frac{2}{k}e^{-\frac{k^2}{12m}}\left( \prod_{j=1}^{\frac{k^2}{4}} \frac{1}{1+\frac{j}{mk^2}}\right)^{-1}\left( \sum_{i=\frac{k^2}{4}+1}^{mk-\frac{k^2}{4}}\prod_{j=1}^i \frac{1}{1+\frac{j}{mk^2}}\right).
    \end{align}
    The second product and the product in the third term can be substituted from \cite{buyukates22ClusterGossip}, hence,
    \begin{align}
        Y &\leq \frac{2\lambda_e}{\lambda}\frac{2}{k}e^{-\frac{k^2}{12m}}e^{\frac{k^2}{16m}}\left( \sum_{i=\frac{k^2}{4}+1}^{mk-\frac{k^2}{4}}e^{-\frac{i^2}{2mk^2}}\right)\\
        &= \frac{2\lambda_e}{\lambda}\frac{2}{k}e^{-\frac{k^2}{48m}}\left( \sum_{i=\frac{k^2}{4}+1}^{mk-\frac{k^2}{4}}e^{-\frac{i^2}{2mk^2}}\right).
    \end{align}
    Using the method in \cite{buyukates22ClusterGossip}, we use step size $\frac{1}{k\sqrt{m}}$, and obtain,
    \begin{align}
        Y &\leq \frac{2\lambda_e}{\lambda}\frac{2}{k}e^{-\frac{k^2}{48m}} k\sqrt{m}\left( \frac{1}{k\sqrt{m}}\sum_{i=\frac{k^2}{4}+1}^{mk-\frac{k^2}{4}}e^{-\frac{i^2}{2mk^2}}\right).\label{Y_sum}
    \end{align}
    
    Now, when converting the summation in \eqref{Y_sum} to a Riemann integral, the upper limit in the summation corresponds to $\infty$ in the upper limit of the integral, since the upper limit is given by $\lim_{k\sqrt{m} \rightarrow \infty} \frac{mk-\frac{k^2}{4}}{k\sqrt{m}} = \lim_{k\sqrt{m}\rightarrow \infty} \sqrt{m} - \frac{k}{4\sqrt{m}} = \infty$. The lower limit can similarly be found to be $\lim_{k\sqrt{m} \rightarrow \infty} \frac{k}{4\sqrt{m}}$, and may go to $\infty$ or to a constant depending on the relation between $k$ and $m$. If $m = o(k^2)$, then the lower limit goes to infinity, although more slowly than the upper limit. Hence, the integral goes to $0$ in this case. On the other hand, if $k = o(\sqrt{m})$, then $e^{-\frac{k^2}{48m}} \rightarrow C$, and we get,
    \begin{align}
        Y \leq \frac{4\lambda_e}{\lambda}C \sqrt{m} \int_{0}^{\infty} e^{-x^2}dx = \sqrt{\frac{\pi}{2}}\frac{4\lambda_e}{\lambda}C \sqrt{m}, \label{grid_Y}
    \end{align}
    where $C$ is a constant. If $k = \sqrt{m}$, then $\sqrt{m} = (mk)^{\frac{1}{3}}$.
    
    Finally, we bound the sum of terms making up $Z$,
    \begin{align}
        Z =& \frac{\lambda_e}{\lambda}+\frac{\lambda_e}{\lambda}\left(1+\sum_{i=mk-\frac{k^2}{4}}^{mk-2} \prod_{j=mk-\frac{k^2}{4}}^i \frac{\lfloor\sqrt{mk-j}\rfloor}{\frac{j}{mk} + \lfloor\sqrt{mk-j-1}\rfloor}\right) \prod_{j=1}^{\frac{k^2}{4}}\frac{\sqrt{j}}{\frac{j}{mk}+ \sqrt{j+1}} \notag\\
        &\times \frac{1}{1-\frac{k}{4m} +  \frac{\sqrt{k}}{2}} \times \left(\prod_{j=\frac{k^2}{4}+1}^{mk-\frac{k^2}{4}-1} \frac{1}{1+\frac{j}{mk^2}}\right)\\
        \leq& \frac{\lambda_e}{\lambda}+\frac{\lambda_e}{\lambda}\left(1+\sum_{i=mk-\frac{k^2}{4}}^{mk-2} \prod_{a=mk-i}^{\frac{k^2}{4}}\frac{\lfloor\sqrt{a}\rfloor}{\frac{mk-a+1}{mk}+ \lfloor\sqrt{a-1}\rfloor}\right) \prod_{j=1}^{\frac{k^2}{4}}\frac{\sqrt{j}}{\sqrt{j+1}} \times \frac{1}{\frac{{\sqrt{k}}}{2}}\\
        \leq& \frac{\lambda_e}{\lambda}+\frac{\lambda_e}{\lambda}{\frac{1}{\frac{\sqrt{k}}{2}}}\frac{2}{k}\left(1+\sum_{i=mk-\frac{k^2}{4}}^{mk-2} \prod_{a=mk-i}^{\frac{k^2}{4}}\frac{\lfloor\sqrt{a}\rfloor}{\lfloor\sqrt{a-1}\rfloor}\right)\\
        \leq&  \frac{\lambda_e}{\lambda}+\frac{\lambda_e}{\lambda}\left(1+\frac{1}{\frac{\sqrt{k}}{2} }\sum_{i=mk-\frac{k^2}{4}}^{mk-2} \frac{1}{\lfloor\sqrt{mk-i-1}\rfloor}\right) \\
        =& \frac{\lambda_e}{\lambda}+\frac{\lambda_e}{\lambda}\left(1+\frac{1}{\frac{\sqrt{k}}{2} }\sum_{l=1}^{\frac{k^2}{4}-1} \frac{1}{\lfloor\sqrt{l}\rfloor}\right)\\
        \leq& \frac{\lambda_e}{\lambda}+\frac{\lambda_e}{\lambda}\left(1+\frac{1}{\frac{\sqrt{k}}{2}}\left(1+\sum_{l=2}^{\frac{k^2}{4}-1} \frac{1}{\sqrt{l}-1}\right)\right)\\
        \leq& \frac{\lambda_e}{\lambda}+\frac{\lambda_e}{\lambda}\left(1+\frac{1}{\frac{\sqrt{k}}{2}}\left(1+\sum_{l=2}^{\frac{k^2}{4}-1} \frac{4}{\sqrt{l}}\right)\right)\\
        \leq& \frac{\lambda_e}{\lambda}+\frac{\lambda_e}{\lambda}\left(1+\frac{1}{\frac{\sqrt{k}}{2}}4k\right)\\
        =& \frac{\lambda_e}{\lambda}\left(2+8\sqrt{k}\right). \label{later_terms}
    \end{align}
    
    Hence, using \eqref{final_eq}, \eqref{grid_Y} and \eqref{later_terms}, we write,
    \begin{align}
        v_1 &\leq X + Y + Z\\
        &\leq \frac{\lambda_e}{\lambda}\left(2 + \beta'(mk)^{\frac{1}{3}}+ 2\sqrt{2\pi}e^{-\frac{k^2}{48m}}\sqrt{m} + 8\sqrt{k}\right).\label{grid_formula}
    \end{align}
    Finally, we conclude that if $k = o(\sqrt{m})$, then $v_1 = O(\sqrt{m})$, otherwise, $v_1 = O((mk)^{\frac{1}{3}})$.
\end{Proof}

\begin{remark}
    We found in Theorem~\ref{thm1} that in a two-dimensional grid network, the version age scaling depends on the relation between the number of nodes in each side of the grid. If $k = o(\sqrt{m})$, then the network behaves like a ring network, its age growing as $O(n^{\frac{1}{2}})$. On the other hand, if $m = o(k^2)$, then the version age grows slowly as $O(n^{\frac{1}{3}})$. Hence, we have a sharp cutoff, $m = k^2$, and drastically different age scaling on either side of this cutoff. To give explicit examples: As intuitively expected, a $2 \times \frac{n}{2}$ grid behaves more like a line (i.e., ring) and has an average version age of $O(n^{\frac{1}{2}})$, while a $\sqrt{n} \times \sqrt{n}$ square grid behaves like a true grid and has an average version age of $O(n^{\frac{1}{3}})$. More interestingly, and somewhat less intuitively, an $n^{\frac{1}{5}} \times  n^{\frac{4}{5}}$ grid behaves like a line (i.e., ring) while a $n^{\frac{2}{5}} \times  n^{\frac{3}{5}}$ grid behaves like a true grid, even though in both cases both dimensions of the grid grow unboundedly with $n$. 
\end{remark}

\section{Version Age in a Generalized Ring}\label{sec5}
In this section, we find an upper bound for the version age of a single node in the generalized ring network, described in Section~\ref{sec2}. We note that the average version age of each node in the network will be the same due to the symmetry of the network. In order to calculate the upper bound, we modify the recursive equations of \cite{yates21gossip} given in \eqref{yates-recursion}, analogous to the method described in Section~\ref{sec4}, but specialized to the topology of the generalized ring this time.

\begin{lemma}\label{lemma4}
For any connected subset $S$ of the generalized ring network, we have,
\begin{align}
    v_S \leq \frac{\frac{\lambda_e}{\lambda}+\frac{|E(S)|}{2f(n)}\max_{i \in N(S)}v_{S \cup \{i\}}}{\frac{|S|}{n}+\frac{|E(S)|}{2f(n)}}.
\end{align}
\end{lemma}

\begin{Proof}
    We start with \eqref{rearranged_recursion} in a similar way to the proof in Lemma~\ref{lemma3}, 
    \begin{align}
       \lambda_e = \lambda_0(S)v_S + \sum_{i \in N(S)}\lambda_i(S)(v_S - v_{S \cup \{i\}}), \label{recur_eq_new}
    \end{align}
    and partition $N(S)$ into $2f(n)$ sets according to the number of incoming nodes into $S$ from any $i \in N(S)$,
    \begin{align}
        A_j = \{i \in N(S): E_S(i) = j\},
    \end{align}
    where $1 \leq j \leq 2f(n)$. Now, we proceed from \eqref{recur_eq_new} as,
    \begin{align}
        \lambda_e &= \lambda_0(S)v_S + \sum_{j=1}^{2f(n)}\sum_{i \in A_j}\lambda_i(S)(v_S - v_{S \cup \{i\}})\\
        &\geq \lambda_0(S)v_S + \sum_{j=1}^{2f(n)}|A_j|\min_{i\in A_j}\lambda_i(S)(v_S - v_{S \cup \{i\}})\\
        &\geq \lambda_0(S)v_S + \sum_{j=1}^{2f(n)}|A_j|\min_{i\in A_j}\lambda_i(S)\min_{i\in A_j}(v_S - v_{S \cup \{i\}})\\
        &= \lambda_0(S)v_S + \sum_{j=1}^{2f(n)}|A_j|\frac{j\lambda}{2f(n)}(v_S - \max_{i\in A_j}v_{S \cup \{i\}})\\
        &\geq \lambda_0(S)v_S + \sum_{j=1}^{2f(n)}|A_j|\frac{j\lambda}{2f(n)}(v_S - \max_{i\in N(S)}v_{S \cup \{i\}})\\
        &= \lambda_0(S)v_S + |E(S)|\frac{\lambda}{2f(n)}(v_S - \max_{i\in N(S)}v_{S \cup \{i\}}),
    \end{align}
    where $|E(S)|=\sum_{j=1}^{2f(n)}j|A_j|$ is the total number of incoming edges into set $S$. Rearranging this with the substitution $\lambda_0(S) = \frac{\lambda|S|}{n}$ proves the lemma.
\end{Proof}

Next, in order to use Lemma~\ref{lemma4}, we need to identify a lower bound for $|E(S)|$ in the case where the number of nodes in $S$ is fixed. To this end, in the following lemma, we find the set of $j$ nodes that has the least number of incoming edges.

\begin{lemma}\label{lemma5}
    Given all connected subsets $S$ such that $|S| = j$, the set which has the minimum number of incoming edges is the continuous set of $j$ nodes. 
\end{lemma}

\begin{figure}
    \centering
    \includegraphics[width = 0.7\linewidth]{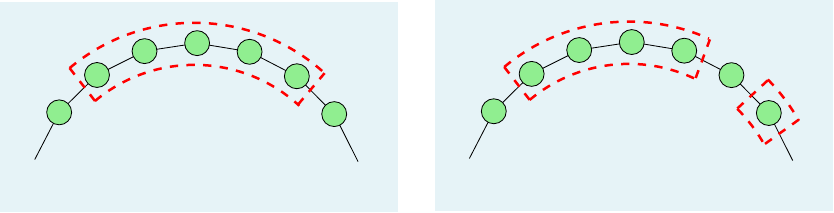}
    \caption{On the left, we have a continuous set of $5$ nodes, marked in red, as described in Lemma~\ref{lemma5}. On the right we have a set of $5$ nodes that is not continuous.}
    \label{fig_ring_contiguous}
\end{figure}

\begin{Proof}
    We define edges that start at a node in $S$ and end at a node in $S$ as inner edges. Let the set of all inner edges in set $S$ be denoted by $\Bar{E}(S)$. We note that the total number of edges incident on nodes in set $S$ is $2jf(n)$, and each edge can only be an inner edge or an incoming edge. Hence, $2|\Bar{E}(S)| + |E(S)| = 2jf(n)$. Thus, showing that the set of continuous nodes has the minimum number of incoming edges is the same as showing that it has the maximum number of inner edges. We will show that $|\Bar{E}(S)|$ is the highest in the continuous set of $j$ nodes.

    Let the set of $j$ continuous nodes be $S_1$, and choose any other connected set of $j$ nodes and call it $S_2$. Next, label each node in both sets as $1_{S_1}, 2_{S_1}. \ldots, j_{S_1}$ and $1_{S_2}, 2_{S_2}. \ldots, j_{S_2}$, respectively. The labels start at the node at one end of the set and end at the other end, covering each node in the order of their position. Now, we compare each $i_{S_1}$ and $i_{S_2}$. We know that both nodes have a total of $2f(n)$ neighbors.

    First, we consider the case where $j \leq f(n)$. In this case, $i_{S_1}$ has all the other nodes in $S_1$ as a neighbor. Hence, it has $j-1$ inner edges associated with it, and this is the highest achievable. $i_{S_2}$ may not have all nodes in $S_2$ as neighbors, and hence has at most the same number of inner edges as $i_{S_1}$. This is true for each consequent node. Hence, adding the number of inner edges of each node in both sets, we see in this case that $S_1$ has more inner edges than $S_2$.

    Next, we consider the case where $f(n) < j \leq 2f(n)$. Suppose the number of nodes in the set is $f(n) + k$, then for $i<k$ and $i>j-k$, $i_{S_1}$ shares an inner edge with $f(n)$ nodes on one side and $i-1$ and $j-i$ neighbors respectively on the other side, which is the highest possible for its position. If $k \leq i \leq j-k$, then all nodes in the set that can share an inner edge with $i_{S_1}$ do so, which again is the highest possible number. Hence, adding the number of inner edges of each node in both sets, we see in this case that $S_1$ has more inner edges than $S_2$.

    Next, we consider the case where $2f(n) < j < n-2f(n)$. If $i<f(n)$, then all $f(n)$ neighbors on one side share an inner edge with $i_{S_1}$, and all $i-1$ neighbors on the other side also share an inner edge with $i_{S_1}$, which leads to $i_{S_1}$ having the highest possible number of inner edges as it shares an inner edge with all possible nodes in its position. Hence, $i_{S_2}$ cannot have more inner edges than $i_{S_1}$. Due to symmetry, this is also true for $n-f(n)<i$. Finally, if $f(n) \leq i \leq n-f(n)$, then all $2f(n)$ neighbors of $i_{S_1}$ share an inner edge with it. Once again, $i_{S_2}$ has at most the same number of inner edges. Hence, in this case, adding up the number of inner edges in order for both sets, we conclude that $S_1$ has more inner edges than $S_2$.

    Finally, if $j \geq n-2f(n)$, then $\mathcal{N}\backslash S$ has the same number of inner edges as $S$. Hence, following the first and second cases, the continuous set has most inner edges.

    From the above four cases, we conclude that the set of continuous nodes has the highest number of inner edges, and hence, the least number of incoming edges. This is shown for a simple example of $|S|=5$ nodes in Fig.~\ref{fig_ring_contiguous}.
\end{Proof}

Now, we state and prove the second main result of this paper in Theorem~\ref{thm3} below.

\begin{theorem}\label{thm3}
    The version age of a single user in the generalized ring network scales as,
    \begin{align}\label{main_result}
        v_1=O\left(\log{f(n)} + \frac{\sqrt{n}}{\sqrt{f(n)}}\right).
    \end{align}
\end{theorem}

\begin{Proof}
    First, using Lemma~\ref{lemma5}, we write the exact lower bounds for incoming edges by counting the number of incoming edges of the sets of continuous nodes. We have three formulae, corresponding to three regions. The regions and the corresponding bounds for incoming edges are:
    \begin{enumerate}
        \item $j \leq f(n): |E(S)| \geq 2jf(n) - j(j-1)$.
        \item $f(n) < j < n-f(n): |E(S)| \geq f(n)(f(n)+1)$.
        \item $n - f(n)  \leq  j: |E(S)| \geq 2(n - j)f(n)  -  (n - j)(n - j - 1)$.
    \end{enumerate}
    We obtain the first bound by counting the number of inner edges for each node, which is $j-1$, and then subtracting it from the total number of neighbors $2f(n)$. Then, the number of incoming edges for each node is $2f(n)-j-1$. Since there are $j$ nodes in total, the number of incoming edges into set $S$ is given by $j(2f(n)-(j-1))$. We can carry out a similar calculation to count the number of edges in the third case. In the second case, we simply calculate the number of incoming edges. The nearest neighbors on each side has $f(n)$ incoming edges, the second nearest neighbor has $f(n)-1$ incoming edges, and so on. Hence, the total number of incoming edges is given by $2 \times \frac{f(n)(f(n)+1)}{2} = f(n)(f(n) + 1)$.

    Now, the recursion can be written using Lemma~\ref{lemma4}. Following the method used to write the recursive equations in Theorem~\ref{thm1}, we define $v_1, v_2, \ldots, v_n = \frac{\lambda_e}{\lambda}$, where $v_1$ is the age of a single node, and each $v_j$ is the age of a set constructed by adding a neighboring node that maximizes the version age among all one-larger sets starting from the set with age $v_{j-1}$. Thus, we analyze the following recursion to find $v_1$, starting at $v_n$,
    \begin{align}
        v_j \leq 
        \begin{cases}
        \dfrac{\frac{\lambda_e}{\lambda} + \frac{2jf(n)-j(j-1)}{2f(n)} v_{j+1}}{\frac{j}{n} + \frac{2jf(n)-j(j-1)}{2f(n)}}, & \quad j \leq f(n),\\
        ~ & ~ \\
        \dfrac{\frac{\lambda_e}{\lambda} + \frac{f(n)+1}{2} v_{j+1}}{\frac{j}{n} + \frac{f(n+1)}{2}}, & \quad f(n) < j < n-f(n),\\
        ~ & ~ \\
        \dfrac{\frac{\lambda_e}{\lambda} + \frac{2(n - j)f(n)  -  (n - j)(n - j - 1)}{2f(n)} v_{j+1}}{\frac{j}{n} + \frac{2(n - j)f(n)  -  (n - j)(n - j - 1)}{2f(n)}}, & \quad n - f(n)  \leq  j.
        \end{cases}
    \end{align}
    
    Now, we write the complete recursion as is done in Theorem~\ref{thm1}, and get a sum containing $n+1$ terms. We divide the sum into three parts corresponding to the terms numbered between $\{1,\ldots,f(n)\}$, $\{f(n) + 1, \ldots, n-f(n)-1\}$ and $\{n-f(n), \ldots, n+1\}$. We call the sum of terms in each of these parts $X$, $Y$, and $Z$, respectively.

    The upper bound for the sum of terms in $X$ can be found as follows,
    \begin{align}
        X &= \frac{\lambda_e}{\lambda}\left(\frac{1}{1+\frac{1}{n}}\right)\left(1+\sum_{i=1}^{f(n)}\prod_{j=1}^i \frac{\frac{2jf(n)-j(j-1)}{2f(n)}}{\frac{j+1}{n}+\frac{2(j+1)f(n)-j(j+1)}{2f(n)}}\right) \label{eq_X_rub}\\
        &\leq \frac{\lambda_e}{\lambda}\left(1+\sum_{i=1}^{f(n)}\prod_{j=1}^i \frac{\frac{2jf(n)-j(j-1)}{2f(n)}}{\frac{2(j+1)f(n)-j(j+1)}{2f(n)}}\right)\\
        &= \frac{\lambda_e}{\lambda}\left(1+\sum_{i=1}^{f(n)}\prod_{j=1}^i \frac{(2f(n)-(j-1))}{2f(n)-j}\frac{j}{j+1}\right)\\
        &= \frac{\lambda_e}{\lambda}\left(1+\sum_{i=1}^{f(n)}\frac{1}{i+1}\frac{2f(n)}{2f(n)-i}\right)\\
        &\leq \frac{\lambda_e}{\lambda}\left(1+\sum_{i=1}^{f(n)}\frac{1}{i}\frac{2f(n)}{2f(n)-i}\right)\\
        &\leq \frac{\lambda_e}{\lambda}\left(2+\sum_{l=1}^{2f(n)-1}\frac{1}{l}\right)\\
        &\leq \frac{\lambda_e}{\lambda}(2+\log{2} + \log{f(n)} + \gamma),
    \end{align}
    where $\gamma \approx 0.577$ is the Euler-Mascheroni constant.
    
    The upper bound for the sum of terms in $Y$ can be found as follows,
    \begin{align}
        Y =& K\frac{\lambda_e}{\lambda}\left(\frac{1}{\frac{f(n)+1}{n}+\frac{f(n)+1}{2}}\right)\left(\frac{f(n)+1}{2}\right)\left(1+\sum_{i=f(n)+2}^{n-f(n)}\prod_{j=f(n)+2}^i \frac{1}{1+\frac{2j}{n(f(n)+1)}}\right), \label{eq_Y_rub}
    \end{align}
    where
    \begin{align}
        K &= \prod_{j=1}^{f(n)}\frac{\frac{2jf(n)-j(j-1)}{2f(n)}}{\frac{j+1}{n} + \frac{2(j+1)f(n)-j(j+1)}{2f(n)}}\\
        &\leq \prod_{j=1}^{f(n)}\frac{j}{j+1}\frac{2f(n)-(j-1)}{2f(n)-j}\\
        &= \frac{2}{f(n)+1}.
    \end{align}
    Substituting this into \eqref{eq_Y_rub}, we get,
    \begin{align}
        Y \leq& \frac{\lambda_e}{\lambda}\frac{2}{f(n)+1}\frac{2}{f(n)+1}\frac{f(n)+1}{2}\left(1+\sum_{i=f(n)+2}^{n-f(n)}\prod_{j=f(n)+2}^i \frac{1}{1+\frac{2j}{n(f(n)+1)}}\right)\\
        =& \frac{\lambda_e}{\lambda}\frac{2}{f(n)+1}\frac{2}{f(n)+1}\frac{f(n)+1}{2}\left(1+\prod_{j=1}^{f(n)+2} \left(1+\frac{2j}{n(f(n)+1)}\right)\sum_{i=f(n)+2}^{n-f(n)}\prod_{j=1}^i \frac{1}{1+\frac{2j}{n(f(n)+1)}}\right)\\
        \leq& \frac{\lambda_e}{\lambda}\frac{2}{f(n)+1}\left(1+\prod_{j=1}^{f(n)+2} \left(1+\frac{2j}{n(f(n)+1)}\right)\sum_{i=1}^{n-f(n)}\prod_{j=1}^i \frac{1}{1+\frac{2j}{n(f(n)+1)}}\right).\label{eq_range_2_2}
    \end{align}
    Next, we take $\log$ of the $i$th product term in the sum of products term,
    \begin{align}
        a_i = \prod_{j=1}^i \frac{1}{1+\frac{2j}{n(f(n)+1)}},
    \end{align}
    which gives us,
    \begin{align}
        -\log{a_i} &= \sum_{j=1}^i \frac{2j}{n(f(n)+1)}\\
        &= \frac{i(i+1)}{n(f(n)+1)},
    \end{align}
    which further gives us,
    \begin{align}
        a_i = e^{-\frac{i(i+1)}{n(f(n)+1)}}.
    \end{align}
    In a similar way, we can conclude that,
    \begin{align}
        \prod_{j=1}^{f(n)+2} \left(1+\frac{2j}{n(f(n)+1)}\right) \approx e^{\frac{f(n)}{n}}.
    \end{align}
    Substituting this into \eqref{eq_range_2_2}, gives,
    \begin{align}
        Y &\leq \frac{\lambda_e}{\lambda}\frac{2}{f(n)+1}\left(1+e^{\frac{f(n)}{n}}\sum_{i=1}^{n-f(n)}e^{-\frac{i(i+1)}{n(f(n)+1)}}\right)\\
        &\leq \frac{\lambda_e}{\lambda}\frac{2}{f(n)+1}\left(1+e^{\frac{f(n)}{n}}\sum_{i=1}^{n}e^{-\frac{i(i+1)}{n(f(n)+1)}}\right)\\
        &\leq \frac{\lambda_e}{\lambda}\frac{2}{f(n)+1}\left(1+e^{\frac{f(n)}{n}}\sum_{i=1}^{n}e^{-\frac{i^2}{n(f(n)+1)}}\right).\label{before_riemann}
    \end{align}
    Now, if $f(n) = o(n)$, then $e^{\frac{f(n)}{n}} \rightarrow 1$, and if $f(n) = \theta(n)$, then $e^{\frac{f(n)}{n}} = C$, where $C$ is a constant.
    Next, we convert the Riemann sum associated with the summation term in \eqref{before_riemann} into a definite integral, and find its exact value. In order to do so, we use step size $\frac{1}{\sqrt{n(f(n)+1)}}$,
    \begin{align}
        \frac{1}{\sqrt{n(f(n)+1)}}\sum_{i=1}^{n}e^{-\frac{i^2}{n(f(n)+1)}} &= \int_0^{\infty} e^{-t^2}dt = \frac{\sqrt{\pi}}{2},
    \end{align}
    as $n \rightarrow \infty$ and $f(n) = o(n)$, and the step size tending to $0$. On the other hand, if $f(n) = \theta(n)$, then the above integral has lower limit $0$ and upper limit a constant, thus resulting in,
    \begin{align}
        \frac{1}{\sqrt{n(f(n)+1)}}\sum_{i=1}^{n}e^{-\frac{i^2}{n(f(n)+1)}} = L,
    \end{align}
    where $L$ is a constant. Using this, we obtain,
    \begin{align}
        \sum_{i=1}^{n}e^{-\frac{i^2}{n(f(n)+1)}} &= L\sqrt{n(f(n)+1)}\\
        &\leq\frac{\sqrt{\pi}}{2}\sqrt{n(f(n)+1)},
    \end{align}
    when $f(n) = \theta(n)$, and
    \begin{align}
        \sum_{i=1}^{n}e^{-\frac{i^2}{n(f(n)+1)}} &= \frac{\sqrt{\pi}}{2}\sqrt{n(f(n)+1)},
    \end{align}
    when $f(n) = o(n)$.
    Substituting this back in \eqref{before_riemann}, we get,
    \begin{align}
        Y &\leq  \frac{\lambda_e}{\lambda}\frac{2}{f(n)+1}\left(1+\frac{\sqrt{\pi}}{2}\sqrt{n(f(n)+1)}\right)\\
        &\approx \sqrt{\pi}\frac{\lambda_e}{\lambda}\frac{\sqrt{n}}{(f(n)+1)^{\frac{1}{2}}}\\
        &\leq \sqrt{\pi}\frac{\lambda_e}{\lambda}\frac{\sqrt{n}}{\sqrt{f(n)}}.
    \end{align}

    The upper bound for the sum of terms in $Z$ can be found as follows,
    \begin{align}
        Z =& 2\frac{\lambda_e}{\lambda} + \frac{\lambda_e}{\lambda}\frac{f(n)+1}{2}\prod_{j=1}^{f(n)+1}\frac{\frac{2jf(n)-j(j-1)}{f(n)}}{\frac{j+1}{n} + \frac{2(j+1)f(n)-j(j+1)}{f(n)}}\times\prod_{j=f(n)+2}^{n-f(n)} \frac{1}{1+\frac{j}{n(f(n)+1)}}\frac{1}{1+\frac{n-f(n)}{n}} \times \notag\\
        &\qquad \qquad \qquad \qquad \quad \times\left(1+ \sum_{i=n-f(n)}^{n-2}\prod_{j=n-f(n)}^i \frac{\frac{(n-j)f(n)-(n-j)(n-j-1)}{f(n)}}{\frac{j+1}{n}+\frac{(n-j-1)f(n)-(n-j-1)(n-j-2)}{f(n)}}\right)\label{eq_Z_rub}\\
        \leq& 3\frac{\lambda_e}{\lambda} + \frac{\lambda_e}{\lambda}\sum_{i=n-f(n)}^{n-2}\prod_{j=n-f(n)}^i   \frac{n-j}{n-j-1}\frac{f(n)-(n-j-1)}{f(n)-(n-j-2)}\label{eq_Z_rub_after}\\
        \leq& 3\frac{\lambda_e}{\lambda} + \frac{\lambda_e}{\lambda}\sum_{i=n-f(n)}^{n-2} \frac{1}{i}\\
        \leq& 3\frac{\lambda_e}{\lambda} + \frac{\lambda_e}{\lambda}\log{f(n)}.
    \end{align}
    In order to go from \eqref{eq_Z_rub} to \eqref{eq_Z_rub_after}, we approximate the first product in the first line of \eqref{eq_Z_rub} following the calculation of $K$ done earlier. We drop the second product in the first line since it is smaller than $1$, and do the regular upper bounding in the second line.
    
    Finally, summing the terms in the three ranges, we obtain,
    \begin{align}
        v_1 \leq& X+Y+Z\\
        \leq& \frac{\lambda_e}{\lambda}(2+\log{2} + \log{f(n)} + \gamma) + \sqrt{\pi}\frac{\lambda_e}{\lambda}\frac{\sqrt{n}}{\sqrt{f(n)}}+ 3\frac{\lambda_e}{\lambda} + \frac{\lambda_e}{\lambda}\log{f(n)}\\
        =& \frac{\lambda_e}{\lambda}(5 + \log{2} + 2\log{f(n)} + \gamma)+\sqrt{\pi}\frac{\lambda_e}{\lambda}\frac{\sqrt{n}}{\sqrt{f(n)}}, \label{eq_ub_ring}
    \end{align}
    giving us the desired result.
\end{Proof}

\subsection{Special Cases of the Generalized Ring Network}\label{sec6}

\subsubsection{Bi-Directional Ring}
We consider the bi-directional ring network in \cite{buyukates21CommunityStructure,buyukates22ClusterGossip}, and it is easy to see that $f(n)=1$. Hence, in this case, we have 
\begin{align}
    v_1 &\leq  \sqrt{\pi}\frac{\lambda_e}{\lambda}\sqrt{n}\\
    &\approx 1.711\frac{\lambda_e}{\lambda}\sqrt{n},
\end{align}
which is in accordance with the result proved in \cite{buyukates21CommunityStructure}.

\subsubsection{Fully-Connected Network}
In this case, it is easy to see that $f(n) = \frac{n-1}{2}$. Substituting this in our equation, we get,
\begin{align}
    v_1 \leq \frac{\lambda_e}{\lambda}(2 + \log{(n-1)}),
\end{align}
which is in accordance with \cite{yates21gossip}, since the exact version age scaling is $\theta(\log{n})$.

\subsubsection{Ring With Fixed Number of Connected Neighbors}

Suppose each node in the ring has $2d$ neighbors, where $d$ is a constant. Then, we can again conclude that, 
\begin{align}
    v_1 \leq \sqrt{\pi}\frac{\lambda_e}{\lambda}\frac{\sqrt{n}}{d^{\frac{3}{2}}}.
\end{align}
Hence, in this case, $v_1 = O(\sqrt{n})$. Thus, being connected to one neighbor on each side or a constant number $d$ neighbors on each side yields the same age scaling on a ring.

\subsubsection{$f(n)=n^\alpha$, $\alpha < 1$}\label{sec_4d}
This set of functions cover functions over a vast range between $f(n) = 1$ and $f(n) = \theta(n)$. We see that the version age in this case scales as a polynomial as,
\begin{align}
    v_1 \leq \sqrt{\pi}\frac{\lambda_e}{\lambda}n^{\frac{1-\alpha}{2}}.
\end{align}
For instance, if $\alpha=\frac{1}{2}$, i.e., each node is connected to $\sqrt{n}$ neighbors on each side on a ring, then the age of a node increases slower than in the case of a regular ring, as $n^\frac{1}{4}$.

\subsubsection{$\frac{n}{\log^2{n}} \leq f(n) < n$}
The version age of a single node grows logarithmically for this set of functions. From \cite{yates21gossip}, we know that for a fully-connected network, the version age scales as $\theta(\log{n})$. Since the networks for these $f(n)$ have smaller number of connections, the version age of a single node in these networks is larger. Hence, a lower bound for the version age is also $\log{n}$. Hence, in this case, we can conclude that the version age of a single user scales as $\theta(\log{n})$.

\begin{remark}\label{remark_generalized_rings}
    In Subsection~\ref{sec6}, we consider the version age scaling for $f(n)=n^\alpha$, and find that the scaling is $O(n^{\frac{1-\alpha}{2}})$ as $n\rightarrow \infty$. Hence, the rational function dominates the logarithm in \eqref{main_result}. However, as $\alpha$ increases, the rational function grows increasingly slowly and dominates the logarithm only at very high values of $n$. In Table~\ref{table_ring_log}, we show the number of nodes in the network required for the rational term to become $10$ times the size of the logarithm. Clearly, this is very high when $\alpha \geq 0.6$. Hence, for all practical purposes, if $f(n) \geq n^{0.6}$, then the version age scaling of a single node in the generalized ring network is logarithmic.
\end{remark}

\begin{table}[h]
    \centering
    \begin{adjustbox}{width = 1\textwidth,center = \textwidth}
    \begin{tabular}{|c|c|c|c|c|c|c|c|c|c|}
    \hline
        $\mathbf{\alpha}$  & 0.1 & 0.2 & 0.3 & 0.4 & 0.5 & 0.6 & 0.7 & 0.8 & 0.9\\
        \hline
        Order of $v_1$ & $n^{0.45}$ & $n^{0.4}$ & $n^{0.35}$ & $n^{0.3}$ & $n^{0.25}$ & $n^{0.2}$ & $n^{0.15}$ & $n^{0.1}$ & $n^{0.05}$\\ 
        \hline
        $n$ & $0$ & $942$ & $24180$ & $955318$ & $1.22 \times  10^8$ & $1.64 \times  10^{11}$ & $3.33 \times  10^{16}$ & $3.9 \times  10^{27}$ & $2.74 \times  10^{63}$\\
        \hline
    \end{tabular}
    \end{adjustbox}
    \vspace{0.2pt}
    \caption{In \eqref{eq_ub_ring}, there are two terms in the upper bound: $\log{f(n)}$ and $\sqrt{n}/\sqrt{f(n)}$. In this table, we list the values of $n$ at which the rational function term begins to dominate the $\log$.}
    \label{table_ring_log}
\end{table}

\section{Version Age in a $m$-Dimensional Hypercube}\label{sec7}
In this section, we find an upper bound for the version age of a single node in a unit $m$-dimensional hypercube, described in Section~\ref{sec2} (see Fig.~\ref{fig_unit_hypercube_network}). Once again, due to the symmetry of the network, the version age of each node in the unit hypercube network will be the same. In order to follow the proof of Theorem~\ref{thm1}, and consequently Lemma~\ref{lemma3}, to find an upper bound for the version age of a single node, we need to find a lower bound for the number of incoming edges. The exact upper bound for the number of inner edges of an induced subgraph of fixed size in a unit hypercube was found in \cite{Hart1976ncube}. Let $h(i)$ be the sum of the binary digits of an integer $i$. Then, the maximum number of inner edges of an induced subgraph $S$ of size $j$ of the hypercube is given by,
\begin{align}
    |\Bar{E}(S)|_{max} = \sum_{i=0}^{j-1}h(i).
\end{align}
Hence, we can find a lower bound for the number of incoming edges using this. We can see that the total number of edges emanating from each node is $m$. Hence, the total number of edges emanating from set $S$ is given by,
\begin{align}
    |E(S)| = mj-2\sum_{i=0}^{j-1}h(i).
\end{align}
Now, we cannot find the exact usable formula for $h(i)$ for each $i$. Hence, we need to find an upper bound for the value of $h(i)$. This upper bound of $h(i)$ was found in \cite{Bush1940sumofdigits}, and is given for $j > 1$ as,
\begin{align}
    \sum_{i=1}^{j-1}h(i) \leq \sum_{i=1}^{\lceil \log_2{(j-1)} \rceil}\frac{1}{2}\left\lfloor \frac{j-1}{2^i} + 1\right\rfloor2^i.
\end{align}
Now, we further upper bound it as follows,
\begin{align}
    \sum_{i=1}^{j-1}h(i) &\leq \left(\sum_{i=1}^{\lceil \log_2{(j-1)} \rceil}\frac{1}{2}\left\lfloor \frac{j-1}{2^i} \right\rfloor2^i \right) + j\\
    &\leq \left(\frac{1}{2}\sum_{i=1}^{\lceil \log_2{(j-1)} \rceil}\left( \frac{j-1}{2^i} \right)2^i \right) + j\\
    &= \left(\frac{1}{2}\sum_{i=1}^{\lceil \log_2{(j-1)} \rceil}(j-1)\right) + j\\
    &\leq \frac{1}{2}j\lceil \log_2{(j)}  \rceil + j\\
    &\approx \frac{1}{2}j\lceil \log_2{(j)}  \rceil.
\end{align}
We can suppress the linear term when compared to the first term as $j \rightarrow \infty$. Hence, we conclude that $|E(S)| \geq mj - j\lceil \log_2{(j)} \rceil$. However, when $j > 2^{m-1}$, the bound simplifies to $|E(S)| \geq 0$, which is true, but impacts our recursive equations. We use the bound $|E(S)| \geq m(n-j) - (n-j)\lceil \log_2{(n-j)}  \rceil$ in this case using the fact that the set with $n-j$ nodes that has the least number of incoming edges has a complement set of $j$ nodes that has the least number of incoming edges among all $j$ sized sets. We now have the following lemma.
\begin{lemma}\label{lemma8}
    Suppose $S$ is an induced subgraph of the hypercube graph and $|S|=j$, $1 \leq j \leq 2^{m-1}$, then we have,
    \begin{align}
        v_S \leq \frac{\frac{\lambda_e}{\lambda} + \frac{j(m-\lceil \log_2{(j)}\rceil)}{m}\max_{i \in S}v_{S \cup \{i\}}}{\frac{j}{2^m}+\frac{j(m-\lceil \log_2{(j)}\rceil)}{m}},
    \end{align}
    and if $j > 2^{m-1}$, then we have,
    \begin{align}
        v_S \leq \frac{\frac{\lambda_e}{\lambda} + \frac{(n-j)(m-\lceil \log_2{(n-j)}\rceil)}{m}\max_{i \in S}v_{S \cup \{i\}}}{\frac{j}{2^m}+\frac{(n-j)(m-\lceil \log_2{(n-j)}\rceil)}{m}}.
    \end{align}
\end{lemma}
The proof of this lemma follows in a similar way to the proof of Lemma~\ref{lemma3}, hence is omitted. 

Next, we state and prove the third main result of this paper in Theorem~\ref{thm4} below.

\begin{theorem}\label{thm4}
    The version age of a single user in the unit hypercube graph grows as $v_1 = O(\log{n}\log{\log{n}})$.
\end{theorem}

\begin{Proof}
The recursion can be written using Lemma~\ref{lemma8}. Following the method used in Theorem~\ref{thm1}, we again define $v_1, v_2, \ldots, v_n = \frac{\lambda_e}{\lambda}$, where $v_1$ is the age of a single node, and each $v_j$ is the age of a set constructed by adding a neighboring node that maximizes the version age among all one-larger sets starting from the set with age $v_{j-1}$. Thus, we analyze the following recursion to find $v_1$, starting at $v_n$,
    \begin{align}
        v_j \leq 
        \begin{cases}
        \dfrac{\frac{\lambda_e}{\lambda} + \frac{j(m-\lceil \log_2{(j)}\rceil)}{m}v_j}{\frac{j}{2^m}+\frac{j(m-\lceil \log_2{(j)}\rceil)}{m}}, & \quad j \leq 2^{m-1},\\
        ~ & ~ \\
        \dfrac{\frac{\lambda_e}{\lambda} + \frac{(n-j)(m-\lceil \log_2{(n-j)}\rceil)}{m}v_j}{\frac{j}{2^m}+\frac{(n-j)(m-\lceil \log_2{(n-j)}\rceil)}{m}}, & \quad 2^{m-1} < j.
        \end{cases}
    \end{align}
    
    Now, we write the complete recursion as is done in Theorem~\ref{thm1}, and get a sum containing $n+1$ terms. We divide the sum into two parts corresponding to the terms numbered between $\{1,\ldots,2^{m-1}\}$ and $\{2^{m-1}+1, \ldots, n+1\}$. We call the sum of terms in each of these parts $X$ and $Y$, respectively. We first find an upper bound for the sum of terms $X$,
    \begin{align}
        X &\leq \frac{\lambda_e}{\lambda}\left(\frac{1}{1+\frac{1}{2^m}}\right)\left(1+\sum_{i=1}^{2^{m-1}-1}\prod_{j=1}^i \frac{\frac{j(m-\lceil \log_2{(j)}\rceil)}{m}}{\frac{j+1}{2^m}+\frac{(j+1)(m-\lceil \log_2{(j+1)}\rceil)}{m}}\right)\\
        &\leq \frac{\lambda_e}{\lambda}\left(1+\sum_{i=1}^{2^{m-1}-1}\prod_{j=1}^i \frac{\frac{j(m-\lceil \log_2{(j)}\rceil)}{m}}{\frac{j+1}{2^m}+\frac{(j+1)(m-\lceil \log_2{(j+1)}\rceil)}{m}}\right)\\
        &\leq \frac{\lambda_e}{\lambda}\left(1+\sum_{i=1}^{2^{m-1}-1}\prod_{j=1}^i \frac{\frac{j(m-\lceil \log_2{(j)}\rceil)}{m}}{\frac{(j+1)(m-\lceil \log_2{(j+1)}\rceil)}{m}}\right)\\
        &\leq \frac{\lambda_e}{\lambda}\left(2+\sum_{i=1}^{2^{m-1}-2}\prod_{j=1}^i \frac{j}{j+1}\frac{m-\lceil \log_2{(j)}\rceil}{m-\lceil \log_2{(j+1)}\rceil}\right)\\
        &= \frac{\lambda_e}{\lambda}\left(2+\sum_{i=1}^{2^{m-1}-2} \frac{1}{i+1}\frac{m}{m-\lceil \log_2{i+1}\rceil}\right)\\
        &= \frac{\lambda_e}{\lambda}\left(2+\sum_{i=2}^{2^{m-1}-2} \frac{1}{i}\frac{m}{m-\lceil \log_2{(i)}\rceil}\right)\\
        &\leq \frac{\lambda_e}{\lambda}\left(2+\sum_{l=1}^{m-2}\sum_{i=2^l}^{2^{l+1}-1} \frac{1}{i}\frac{m}{m-l}\right)\\
        &= \frac{\lambda_e}{\lambda}\left(2+\sum_{l=1}^{m-2}\frac{m}{m-l}\sum_{i=2^l}^{2^{l+1}-1} \frac{1}{i}\right)\\
        &= \frac{\lambda_e}{\lambda}\left(2+\log{2}\sum_{l=1}^{m-2}\frac{m}{m-l}(\log{(2^{l+1}-1)}-l)\right)\\
        &\leq \frac{\lambda_e}{\lambda}\left(2+\log{2}\sum_{l=1}^{m-2}\frac{m}{m-l}((l+1)-l)\right)\\
        &\leq \frac{\lambda_e}{\lambda}\left(2+\log{2}\sum_{l=1}^{m-2}\frac{m}{m-l}\right)\\
        &\leq \frac{\lambda_e}{\lambda}\left(2+\log{(2)}m\log{m}\right).
    \end{align}
    
    Next, we find an upper bound for the sum of terms $Y$,
    \begin{align}
        Y \leq& \frac{\lambda_e}{\lambda} \left(\prod_{j=1}^{2^{m-1}-1}\frac{\frac{j(m-\lceil \log_2{(j)}\rceil)}{m}}{\frac{j+1}{2^m}+\frac{(j+1)(m-\lceil \log_2{(j+1)}\rceil)}{m}}\right)\frac{2^{m-1}}{m}\frac{1}{\frac{2^{m-1}+1}{2^m} + \frac{2^{m-1}+1}{m}}\\
        &\times\left(1+\sum_{i=2^{m-1}+1}^{2^m-1}\prod_{j=2^{m-1}+1}^i \frac{\frac{(n-j)(m-\lceil \log_2{(n-j)}\rceil)}{m}}{\frac{j+1}{2^m}+\frac{(n-j-1)(m-\lceil \log_2{(n-j-1)}\rceil)}{m}}\right)\\
        \leq& \frac{\lambda_e}{\lambda}\left(\prod_{j=1}^{2^{m-1}-1}\frac{\frac{j(m-\lceil \log_2{(j)}\rceil)}{m}}{\frac{(j+1)(m-\lceil \log_2{(j+1)}\rceil)}{m}}\right)\left(1+\sum_{i=2^{m-1}+1}^{2^m-1}\prod_{j=2^{m-1}+1}^i \frac{\frac{(n-j)(m-\lceil \log_2{(n-j)}\rceil)}{m}}{\frac{(n-j-1)(m-\lceil \log_2{(n-j-1)}\rceil)}{m}}\right)\\
        \leq& \frac{\lambda_e}{\lambda} \frac{m}{2^{m-1}}\left(1+\sum_{i=2^{m-1}+1}^{2^m-1}\prod_{j=2^{m-1}+1}^i \frac{n-j}{n-j-1}\frac{m-\lceil \log_2{(n-j)}\rceil}{m-\lceil \log_2{(n-j-1)}\rceil}\right)\\
        =& \frac{\lambda_e}{\lambda}\left(1+\frac{m}{2^{m-1}}\sum_{i=2^{m-1}+1}^{2^m-1} \frac{n-i}{2^{m-1}-2}(m-\lceil \log_2{(n-i)}\rceil)\right)\\
        =& \frac{\lambda_e}{\lambda}\left(1+\frac{m}{(2^{m-1})(2^{m-1}-2)}\sum_{l=1}^{2^{m-1}-1} (2^{m-1}-l)(m-\lceil \log_2{(2^{m-1}-l)}\rceil)\right)\\
        =& \frac{\lambda_e}{\lambda}\left(1+\frac{m}{(2^{m-1})(2^{m-1}-2)}\sum_{a=1}^{2^{m-1}-1} a(m-\lceil \log_2{(a)}\rceil)\right)\\
        \leq& \frac{\lambda_e}{\lambda}\left(1+\frac{m}{(2^{m-1})(2^{m-1}-2)}\left(3m+\sum_{b = 1}^{m-2}\sum_{a=2^b+1}^{2^b} a(m-\lceil \log_2{(a)}\rceil)\right)\right)\\
        =& \frac{\lambda_e}{\lambda}\left(1+\frac{m}{(2^{m-1})(2^{m-1}-2)}\left(3m+\sum_{b = 1}^{m-2}(m-(b+1))\sum_{a=2^b+1}^{2^b} a\right)\right)\\
        \approx& \frac{\lambda_e}{\lambda}\left(1+\frac{m}{(2^{m-1})(2^{m-1}-2)}\left(3m+3\sum_{b = 1}^{m-2}(m-(b+1))2^{2b-1}\right)\right)\\
        \leq& \frac{\lambda_e}{\lambda}\left(1+\frac{m}{(2^{m-1})(2^{m-1}-2)}\left(3m+3\sum_{t = 1}^{m-2}t2^{2m-2t-2}\right)\right)\\
        =& \frac{\lambda_e}{\lambda}\left(1+\frac{m}{(2^{m-1})(2^{m-1}-2)}\left(3m+3.2^{2m}\sum_{t = 1}^{m-2}\frac{t}{2^{2t+2}}\right)\right)\\
        =& \frac{\lambda_e}{\lambda}\left(1+\frac{m}{(2^{m-1})(2^{m-1}-2)}\left(3m+3.2^{2m}(\frac{1}{9}\frac{1}{4^{m-1}}(8+4^m-12m)))\right)\right)\\
        \approx& \frac{\lambda_e}{\lambda} \left(1 + \frac{16}{3}m\right).
    \end{align}
    
    Finally, we use $X$ and $Y$ to upper bound $v_1$, and also substitute $m = \log_2{n}$,
    \begin{align}
        v_1 &\leq X+Y\\
        &\leq \frac{\lambda_e}{\lambda}(3 + \frac{16}{3}m + \log{(2)}m\log{m})\\
        &= \frac{\lambda_e}{\lambda}(3 + \frac{16}{3}\log_2{(n)} + \log{(2)}\log_2{(n)}\log_2{\log_2{(n)}})\\
        &\leq \frac{\lambda_e}{\lambda}C\log{n}\log{\log{n}},
    \end{align}
    where $C$ is a constant.
\end{Proof}

\section{Discussion and Remarks}\label{sec9}

In this section, we discuss implications and possible extensions of our results via a sequence of remarks. We start with a general remark, and then present our remarks grouped by each network structure in the following subsections. 

\begin{remark}
We note that there are several works in the literature which consider gossip networks with the version age metric in different settings. These include: adversarial models such as timestomping \cite{kaswan22timestomp} and jamming \cite{kaswan22jamming}, hierarchical structures such as a network with a community structure \cite{buyukates21CommunityStructure}, extended gossip algorithms such as file slicing and network coding \cite{kaswan22slicingcoding}, and opportunistic gossiping in multiple access and decentralized networks \cite{mitra_allerton22, mitra2023timely}. All these papers use either the ring network or the fully-connected network as the network model. With the results of the present paper,  extensions of the results in the existing literature to the generalized networks, such as grids, generalized rings and hypercubes, are possible.  
\end{remark}

\subsection{Remarks on the Grid Network}

\begin{remark}\label{remark1}
    In \eqref{beta}, $\beta = (\frac{2}{3})^{2/3}\Gamma(\frac{1}{3})$, where $\Gamma(\cdot)$ is the gamma function. Thus, $\beta' = e^{-\frac{\gamma}{2}}\times e^{\pi^2/48} \times (\frac{2}{3})^{2/3}\Gamma(\frac{1}{3}) \approx 1.882$. Hence, if $m = o(k^2)$, then as $n \rightarrow \infty$, the second term in \eqref{grid_formula} dominates, and the version age of a single node is upper bounded as,
    \begin{align}
        v_1 \leq 3.764\frac{\lambda_e}{\lambda}n^{\frac{1}{3}}. \label{54}
    \end{align}
    On the other hand, if $k = o(\sqrt{m})$, then the third term dominates in \eqref{grid_formula}. The constant multiplier depends on the relation between $k$ and $m$ in this case.
\end{remark}

\begin{remark}\label{remark2}
    In a square two dimensional grid, $k=m$, and hence the version age scaling is given by $v_1 = O(n^{\frac{1}{3}})$. Moreover, if a square two dimensional grid is such that there are no wrap-around edges, then the version age still has the same order. This is because, the spiral still is the best up to a point, and then the next best set makes use of three sides of the grid, leaving only one side for incoming edges. In this new case, the number of incoming edges is lower bounded by $\sqrt{n}$. Hence, $|E(S)|$ is bounded below by $\sqrt{j}$, until we reach the inverse spiral set. Thus, the version age still is bounded by $Kn^{\frac{1}{3}}$. Therefore, the order does not change in this case. Hypothetically, this is because the deletion of the wrapped around edges does not change the order of the diameter of the network, which is still $\theta(\sqrt{n})$.
\end{remark}

\begin{figure}[t]
    \centering
    \includegraphics[width = 0.7\linewidth]{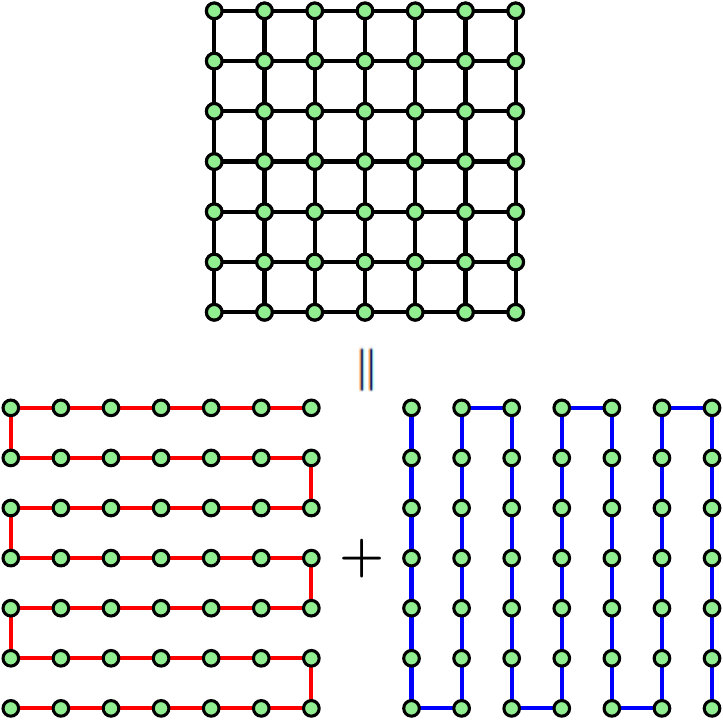}
    \caption{The grid can be thought of as two conjoined line networks, which reduces the version age from $O(n^{\frac{1}{2}})$ to $O(n^{\frac{1}{3}})$.}
    \label{fig_conjoined_grid}
\end{figure}
    
\begin{remark}\label{remark4}
    It is well known that conjoining two networks disseminating the same information speeds up the information diffusion \cite{yagan2013conjoining}. For example, suppose that there is a group of friends living in the same city, who meet each other regularly. They want to be up-to-date about the happenings in each other's lives. Then, we see that they will receive updates about their friends faster if they were connected to each other on a social media platform and also met each other in-person, rather than doing just one of the two. This is an interesting perspective with which we can also look at the grid network as a conjoined network of two lines (i.e., two rings) as shown in Fig.~\ref{fig_conjoined_grid}, which improves the version age from $O(n^{\frac{1}{2}})$ to $O(n^{\frac{1}{3}})$. In Fig.~\ref{fig_conjoined_grid}, we view the grid in black as a conjoined network of a red line and a blue line.    
\end{remark}

\begin{remark}\label{remark5}
    Reference \cite{buyukates21CommunityStructure} considers gossip networks with a community structure. It considers a system where a source node receives an update and disseminates it to cluster heads, which in turn disseminate this information to cluster nodes under them. It is shown in \cite{buyukates21CommunityStructure} that having a community structure improves the version age for several community structures. These community (i.e., cluster) structures include the disconnected network, the ring network and the fully-connected network. Following Section~\ref{sec3}, we can upper bound the recursion found in \cite[Thm.~1]{buyukates21CommunityStructure} as,
    \begin{align}
        v_S \leq &\frac{\lambda_e + \lambda_c(S)v_c}{\lambda_c(S) + |N(S|\min_{i \in N(S)}\lambda_i(S)}+ \frac{|N(S)|\min_{i \in N(S)}\lambda_i(S)\max_{i \in N(S)}v_{S \cup \{i\}}}{\lambda_c(S) + |N(S|\min_{i \in N(S)}\lambda_i(S)},
    \end{align}
    where there are $x$ communities each with $y$ nodes such that $n = xy$ and $v_c = x\lambda_e/\lambda_s$. Then, assuming $\lambda_c = \lambda$ and that clusters are square grid networks as we have defined in this paper, we can upper bound $v_1$ in one cluster of the community as,
    \begin{align}
        v_1 \leq v_c(1 - a_{y}) + 2\beta'\frac{\lambda_e}{\lambda}y^{\frac{1}{3}},
    \end{align}
    where $a_y$ is as defined in \eqref{def_ak}. Now, as $n \rightarrow \infty$, $a_y \rightarrow 0$. Hence, we have,
    \begin{align}
        v_1 \leq x\frac{\lambda_e}{\lambda_s} + 2\beta'\frac{\lambda_e}{\lambda}y^{\frac{1}{3}}.
    \end{align}
    Choosing $x = n^\frac{1}{4}$ and $y = n^\frac{3}{4}$, by adding community structure, we can improve the version age from $O(n^{\frac{1}{3}})$ to $O(n^{\frac{1}{4}})$.
\end{remark}

\subsection{Remarks on the Generalized Ring}

\begin{remark}\label{remark6}
    Extremal animals have been a topic of study in graph theory for a long time\cite{harary1976extremal}. These are connected subgraphs with minimum or maximum number of neighboring nodes, edges or faces in a graph. In this context, Lemma~\ref{lemma5} finds the minimal edge animal for the generalized ring network. 
\end{remark}

\begin{remark}\label{remark7}
    In Section~\ref{sec4}, it was shown that a two-dimensional grid has version age scaling of $O(n^{\frac{1}{3}})$. Each node in the grid network has $4$ neighbors. However, in order to achieve a version age scaling of $O(n^{\frac{1}{3}})$ in the generalized ring network, we need $f(n) = n^{\frac{1}{3}}$, i.e. each node needs $2n^{\frac{1}{3}}$ neighbors. One way to explain this difference in the requirement for connectivity to achieve the same version age scaling is the following: According to Remark~\ref{remark4}, we can view the grid as a ring network with $n$ connections which are not local in nature. Hence, although the number of connections in a grid network is much fewer compared to the generalized ring network, the version age scaling is the same. This shows that the geometry of a network can affect the version age significantly, and having few connections between nodes far away is better than having relatively dense connections which are local in nature.
\end{remark}

\subsection{Remarks on the Unit Hypercube}

\begin{remark}\label{remark9}
    We note that the version age scaling in this case is very close to the fully-connected network. In order for the version age of a generalized ring network to have the same scaling, we need $f(n) = \frac{n}{(\log{n}\log{\log{n}})^2}$ connections on each side. This graph has diameter $(\log{n}\log{\log{n}})^2$. This is much higher than the diameter of the unit hypercube, which is $\log{n}$. Moreover, each node in the generalized ring network has $\frac{2n}{(\log{n}\log{\log{n}})^2}$ neighbors. This is very large when compared to a node in the unit hypercube network, where each node has $\log_2 n$ neighbors. Hence, we notice that the geometry of a graph is very important while determining the scaling of the average version age.
\end{remark}

\begin{remark}\label{remark10}
    In Fig.~\ref{fig_unit_hypercube}, the simulated version age of the unit hypercube follows the logarithm closely, when compared to the upper bound found in Section~\ref{sec7}. Moreover, in order for the upper bound to dominate the logarithm, i.e., be $10$ times larger than the logarithm, we need $e^{e^{10}}$ nodes, which implies that the version age of a single node in the hypercube is logarithmic for all practical purposes.
\end{remark}

\subsection{Remarks on General Hypercubes}

\begin{remark}\label{remark11}
In this remark, we consider the problem of finding the version age of a single node in a general hypercube graph as defined in Section~\ref{sec2}. It is clear that due to the symmetry of the network, the version ages of any two nodes in the graph are equal. In order to write an upper bound recursion for this problem, we need to find a lower bound for $|E(S)|$.

For an infinite grid graph $\mathbb{Z}^d$, it is found in \cite{Agnarsson2013ExtremalSO} that the number of inner edges $|\Bar{E}(S)|$ is upper bounded as follows (assuming $|S| = j$),
\begin{align}
    |\Bar{E}(S)| &\leq \lfloor jd(1-j^{-\frac{1}{d}}) \rfloor\\
    &\leq jd - \lfloor dj^{\frac{d-1}{d}} \rfloor\\
    &\leq jd - d\lfloor j^{\frac{d-1}{d}} \rfloor,
\end{align}
and since the total number of edges is given by $2jd$, the number of incoming edges can be lower bounded as,
\begin{align}\label{eq_d_dim_ub}
    |E(S)| \geq 2d\lfloor j^{\frac{d-1}{d}} \rfloor.
\end{align}
This bound is an extension to the one found for the two-dimensional grid that we considered in Section~\ref{sec4}. The bound in \eqref{eq_d_dim_ub} is true for the case where the grid graph is infinite. However, in our case, the graph is finite (and wrapped-around at the boundaries). Hence, we need to consider the boundary effects, which will lead to different lower bounds for the number of incoming edges depending on how many boundary edges are used to decrease the number of incoming edges. Moreover, each set can have at most $2d$ faces. Hence, we conjecture that the number of incoming edges is greater than $j^{\frac{d-1}{d}}$ in the finite $d$-dimensional grid, up to $j = \frac{n}{2}$. For $j > \frac{n}{2}$, we can see that the complement set of the best set for $n-j$ is the set with the least number of incoming edges, and hence the bound $j^{\frac{d-1}{d}}$ works in this case as well. 

Now, we can write the upper bound recursive equations following the proof of Lemma~\ref{lemma3}, and construct sets $v_1,v_2,\ldots,v_n = \frac{\lambda_e}{\lambda}$ as in Theorem~\ref{thm1} to get the following recursion,
\begin{align}
    v_j \leq 
        \begin{cases}
        \dfrac{\frac{2d\lambda_e}{\lambda} + j^{\frac{d-1}{d}}v_j}{\frac{j}{n}+j^{\frac{d-1}{d}}}, & \quad j \leq \frac{n}{2},\\
        ~ & ~ \\
        \dfrac{\frac{2d\lambda_e}{\lambda} + (n-j)^{\frac{d-1}{d}}v_j}{\frac{j}{n}+(n-j)^{\frac{d-1}{d}}}, & \quad \frac{n}{2} < j.
        \end{cases}
\end{align}

We then write out the complete recursion as done in Theorem~\ref{thm1} and get $n+1$ terms. We divide these terms into two parts, corresponding to the ranges $\{1,\ldots,\frac{n}{2}\}$ and $\{\frac{n}{2}+1,\ldots,n+1\}$. We have two sum of terms corresponding to these ranges for $j$, say $X$ and $Y$. We find the upper bound for $X$ as follows,
\begin{align}
    X &\leq \frac{2d\lambda_e}{\lambda}\left(\frac{1}{1+\frac{1}{n}}\right)\left(1 + \sum_{i=1}^{\frac{n}{2}} \prod_{j=1}^i \frac{j^\frac{d-1}{d}}{ (j+1)^{\frac{d-1}{d}} + \frac{j+1}{n}}\right)\\ 
    &\leq \frac{2d\lambda_e}{\lambda}\left(\frac{1}{1+\frac{1}{n}}\right)\left(1 + \sum_{i=1}^n \prod_{j=1}^i \frac{j^\frac{d-1}{d}}{(j+1)^{\frac{d-1}{d}} + \frac{j}{n}}\right)\\ 
    &\leq \frac{2d\lambda_e}{\lambda}\left(1 + \sum_{i=1}^n \prod_{j=1}^i \frac{j^\frac{d-1}{d}}{(j+1)^{\frac{d-1}{d}} + \frac{j}{n}}\right),
\end{align}
as $n \rightarrow \infty$. Next, we find the order of the right hand side,
\begin{align}
    X &\leq \frac{2d\lambda_e}{\lambda}\left(1 + \sum_{i=1}^n \prod_{j=1}^i \frac{1}{(1+\frac{1}{j})^{\frac{d-1}{d}} + \frac{j^{\frac{1}{d}}}{n}}\right) \label{eq_binom_approx_before_d}\\ 
    &\leq \frac{2d\lambda_e}{\lambda}\left(1 + \sum_{i=1}^n \prod_{j=1}^i \frac{1}{1 + \frac{d-1}{d}\frac{1}{j} - \frac{d-1}{2d^2}\frac{1}{j^2} + \frac{j^{\frac{1}{d}}}{n}}\right),\label{eq_binom_approx_after_d}
\end{align}
where \eqref{eq_binom_approx_after_d} uses the approximate binomial expansion to bound \eqref{eq_binom_approx_before_d}. Now, we define,
\begin{align}
    a_i &= \prod_{j=1}^i \frac{1}{1 + \frac{d-1}{d}\frac{1}{j} - \frac{d-1}{2d^2}\frac{1}{j^2} + \frac{j^{\frac{1}{d}}}{n}}.
\end{align}
Then, we have,
\begin{align}
    -\log{a_i} &= \sum_{j=1}^i \log\left(1+\frac{d-1}{d}\frac{1}{j} - \frac{d-1}{2d^2}\frac{1}{j^2} + \frac{j^{\frac{1}{d}}}{n}\right)\\
    &\approx \sum_{j=1}^i \left(\frac{d-1}{d}\frac{1}{j} - \frac{d-1}{2d^2}\frac{1}{j^2} + \frac{j^{\frac{1}{d}}}{n}\right)\\
    &\approx \frac{d-1}{d}\log{i} - C + \frac{di^{\frac{d+1}{d}}}{(d+1)n},
\end{align}
where $C$ is a finite positive constant upper bounded by $(d-1)\pi^2/6d^2$, and the third term is bounded by the first term of the Euler-MacLaurin series expansion. Substituting this back in \eqref{eq_approx}, we obtain,
\begin{align}
    X &\leq \frac{2d\lambda_e}{\lambda}\left(1 + k\sum_{i=1}^n e^{-(\frac{d-1}{d})\log{i} - \frac{di^{\frac{d+1}{d}}}{(d+1)n}}\right)\\
    &= \frac{2d\lambda_e}{\lambda}\left(1 + k\sum_{i=1}^n \frac{1}{i^{\frac{d-1}{d}}} e^{-\frac{di^{\frac{d+1}{d}}}{(d+1)n}}\right), \label{eq_reimann_sum_d}
\end{align}
where $k = e^{(d-1)\pi^2/6d^2}$. We define a function $f$ and then write the Riemann sum with step size $n^{d/d+1}$ and construct an integral as follows,
\begin{align}
    f(x) &= \frac{1}{x^{\frac{d-1}{d}}}e^{-(\frac{d}{d+1})x^{\frac{d+1}{d}}}\\
    \sum_{i=1}^n \frac{1}{n^{\frac{d}{d+1}}} f\left(\frac{i}{n^{\frac{d}{d+1}}}\right) &= \int_{0}^{\infty} \frac{1}{t^{\frac{d-1}{d}}} e^{-(\frac{d}{d+1})t^{\frac{d+1}{d}}}dt = L < \infty.
\end{align}
Hence, we can continue as follows,
\begin{align}
    \sum_{i=1}^n \frac{1}{n^{\frac{d}{d+1}}} f\left(\frac{i}{n^{\frac{d}{d+1}}}\right) &= \frac{1}{n^{1/(d+1)}} \sum_{i=1}^n \frac{1}{i^{\frac{d-1}{d}}} e^{-\frac{di^{\frac{d+1}{d}}}{(d+1)n}} = L.
\end{align}
Finally, continuing from \eqref{eq_reimann_sum_d}, we have,
\begin{align}
    X &\leq \frac{2d\lambda_e}{\lambda}\left(1 + k\sum_{i=1}^n \frac{1}{i^{\frac{d-1}{d}}} e^{-\frac{di^{\frac{d+1}{d}}}{(d+1)n}}\right)\\
    &= \frac{2d\lambda_e}{\lambda}(1 + kLn^{1/(d+1)}).
\end{align}
Next, we write the recursion for $Y$ in the same way we did for $Z$ in Theorem~\ref{thm1} and observe that $Y = O(\log{n})$.

Hence, we conclude that $v_1 = O(n^{1/(d+1)})$, if our conjecture for the lower bound for the number of incoming edges is correct. We note that the formula $v_1 = O(n^{1/(d+1)})$ satisfies the results for $d=0, 1, 2$, where the version age of a single node in these cases scales as $O(n)$, $O(n^{\frac{1}{2}})$ and $O(n^{\frac{1}{3}})$, respectively, in a disconnected network, a line network and a square two-dimensional grid, which are hypercube networks with $d=0$, $d=1$ and $d=2$, respectively. 
\end{remark}

\begin{remark}\label{remark13}
    Even though these results show the improvement in version age as the connectivity increases, these results still do not describe the entire spectrum of connectivity going from a disconnected network where the version age scales as $O(n)$ all the way to the fully-connected network where the version age scales as $O(\log n)$, since the nodes in the $d$-dimensional grid have $2d$ neighbors, which is still a constant even though it increases with $d$. In contrast, in the fully-connected network, each node has $n-1$ neighbors, which depends on the number of nodes $n$ in the network.
\end{remark}

\begin{remark}
    We can carry out a similar exercise as the one carried out in Remark~\ref{remark5} for a $d$-dimensional grid and conclude that we can improve the version age from $O(n^{\frac{1}{d+1}})$ to $O(n^{\frac{1}{d+2}})$ by introducing community structure to the network.
\end{remark}

\section{Numerical Results} \label{sec10}
In this section, we provide our numerical results grouped by each network structure studied in this paper, in the following subsections. 

\begin{figure}[t]
        \centering
        \includegraphics[width = 0.7\textwidth]{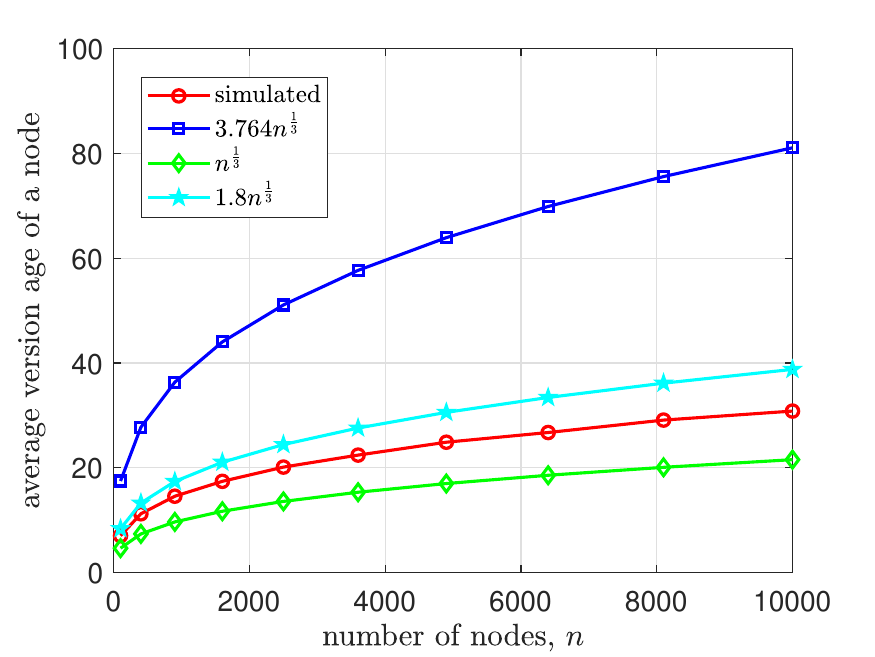}
        \caption{Version age of a single node versus the number of nodes in a two-dimensional grid gossip network.}
        \label{fig_grid_nxn}    
\end{figure}

\subsection{Numerical Results for the Grid Network}
We perform numerical simulations to find the version age of a single node in the two-dimensional grid network, and compare them to the bounds we obtained. We set $\lambda_e = 1$, $\lambda = 1$ in each case. First, we simulate the square two-dimensional grid by varying $n$ from $100$ nodes corresponding to a $10 \times 10$ grid, to $10,000$ nodes corresponding to a $100 \times 100$ grid, such that the number of nodes on one edge of the grid increases by increments of $10$. We plot the results in Fig.~\ref{fig_grid_nxn}. The simulated version age is plotted along with the upper bound $3.764n^{\frac{1}{3}}$ we obtained in \eqref{54}. We see that our calculated upper bound bounds the simulated version age. We also plot $1.8n^{\frac{1}{3}}$ and $n^{\frac{1}{3}}$ to provide closer upper and lower bounds for the simulated version age. By trying various multiples for $n^{\frac{1}{3}}$ to match with the simulated version age, we find that $1.45n^{\frac{1}{3}}$ is a close approximation to the simulated version age we have obtained. 

\begin{figure}[t]
    \centering
    \begin{subfigure}{0.48\textwidth}
        \centering
        \includegraphics[width = \textwidth]{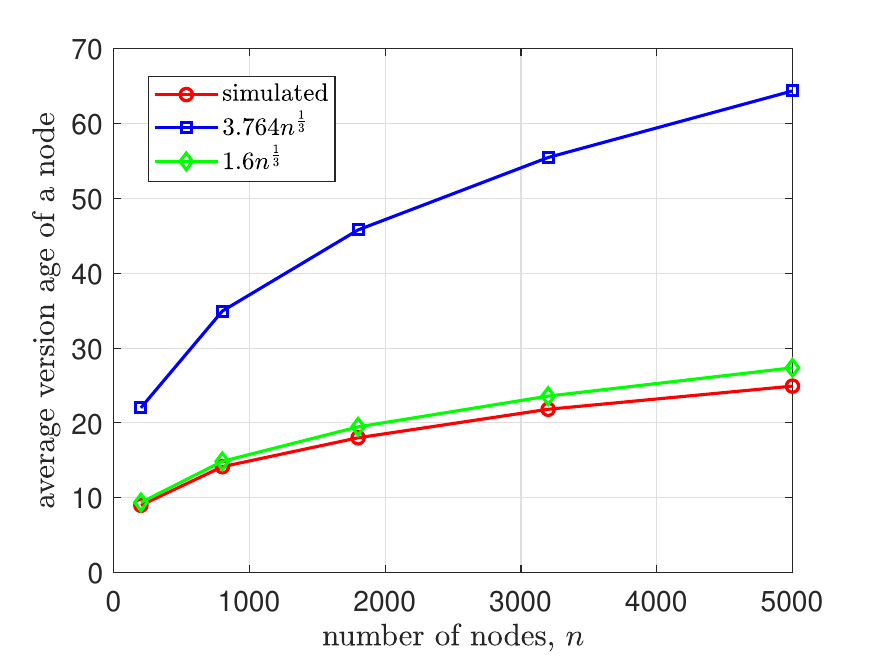}
        \caption{}
        \label{fig_grid_mxn}
    \end{subfigure}
    \hfill
    \begin{subfigure}{0.48\textwidth}
        \centering
        \includegraphics[width = \textwidth]{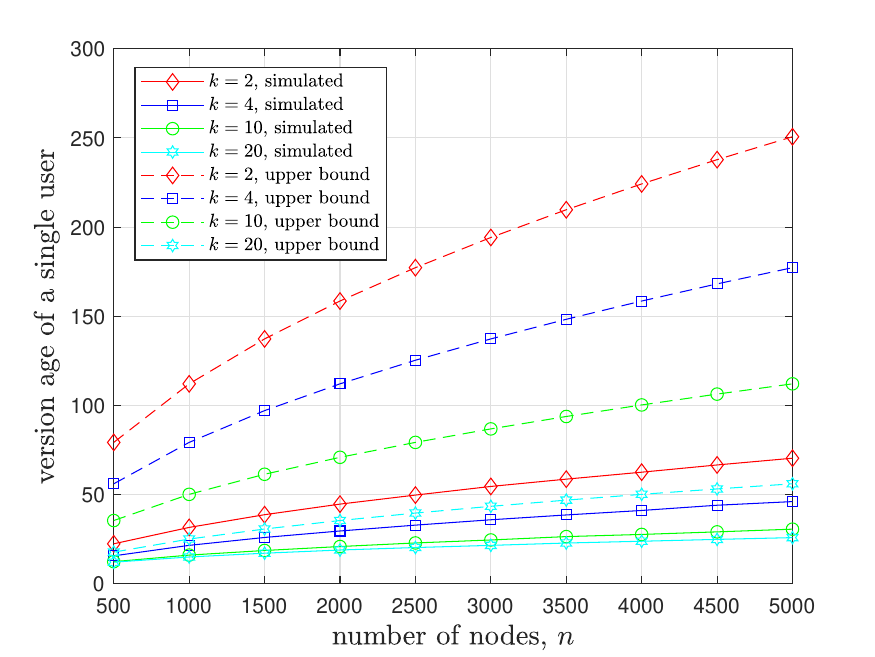}
        \caption{}
        \label{fig_grid_square_root}
    \end{subfigure}
    \caption{(a) Variation of the version age of a single node in the grid network where one side is half the length of the other side, i.e., $m=2k$, and the number of nodes $n$, i.e., $n=mk$. (b) Variation of the version age of a single node in the grid network where one side is constant, and the other side is increasing, and the number of nodes $n$.}
\end{figure}

We also simulate a rectangular network where $m=2k$, and the larger side is varied from $20$ to $100$ in steps of $20$. We plot the results in Fig.~\ref{fig_grid_mxn}. We see that the upper bound we found still works in this case, since we are in the regime where $m = o(k^2)$. Finally, we simulate a network where one side of the grid is constant. We choose the values such that the constant length side is the smaller one in each case, i.e., we choose $k$ to be constant. The number of nodes varies from $500$ to $5000$ with a step size of $500$. We consider four values for $k$, specifically $2$, $4$, $10$ and $20$. We plot the results in Fig.~\ref{fig_grid_square_root}. We note that the upper bound is given by $2\sqrt{2\pi}\sqrt{\frac{n}{k}}$ in this case. Hence, as the value of $k$ increases, the bound becomes smaller, along with the simulated version age. This is because when $k$ is larger, the grid is more densely connected, resulting in a smaller version age.

\begin{figure}[th!]
    \centering
    \begin{subfigure}{0.48\textwidth}
        \centering
        \includegraphics[width = \textwidth]{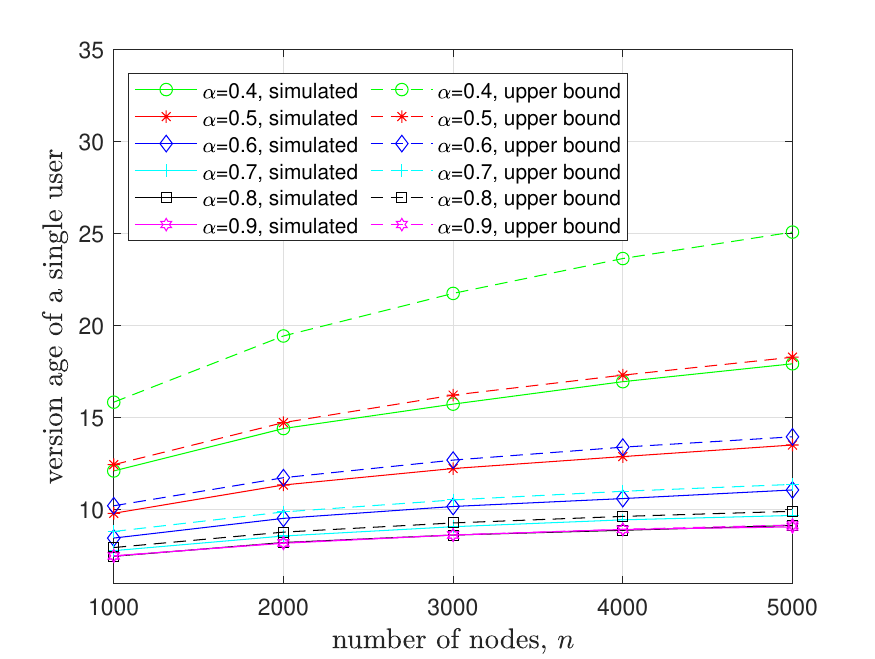}
        \caption{}
        \label{fig_ring_no_log}
    \end{subfigure}
    \hfill
    \begin{subfigure}{0.48\textwidth}
        \centering
        \includegraphics[width = \textwidth]{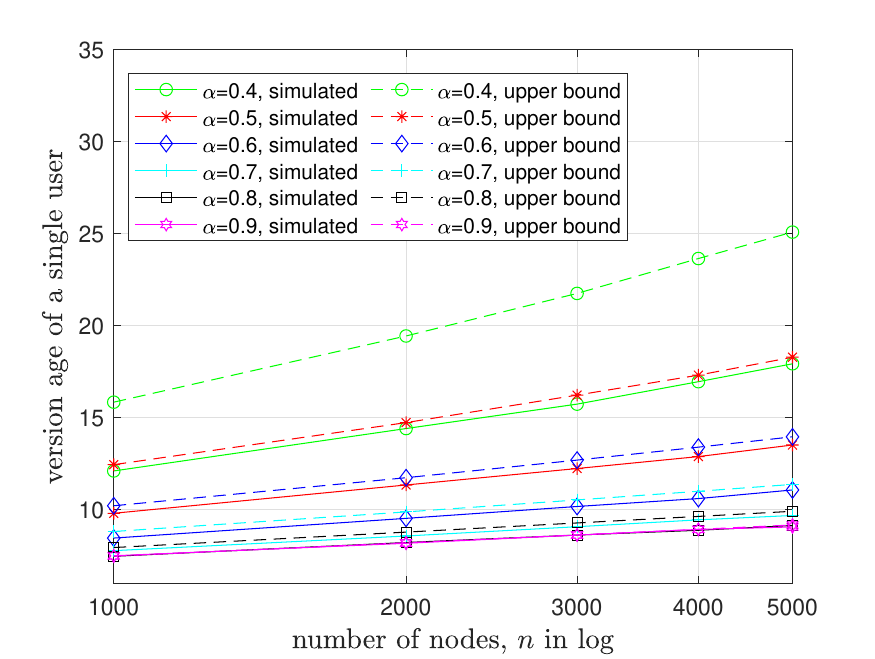}
        \caption{}
        \label{fig_ring_log}
    \end{subfigure}
    \caption{We plot the simulated version age for the generalized ring networks along with the upper bound in both linear and log scale. Here $f(n) = n^\alpha$. The left figure has a linear x-axis and the right figure has a logarithmic x-axis.} 
\end{figure}

\begin{table}[th!]
\begin{subtable}{1\textwidth}
\centering
   \begin{tabular}{|l|c|c|c|c|c|}
      \hline
      {\(\;\;\;\;\;\;\;\;\;\;\;\;\;\;\;\;n\)} & {\(10^4\)} & {\(10^5\)} & {\(10^6\)} & {\(10^7\)} & {\(10^8\)}\\ 
      \hline
      Recursive upper bound & 124.641 & 395.658 & 1252.645 & 3962.659 & 12532.474\\ 
      \hline
      Upper bound from \eqref{eq_ub_ring} & 183.515 & 566.769 & 1778.723 & 5611.261 & 17730.808\\
      \hline
   \end{tabular}
   \caption{Version age of a single node when $\alpha = 0$.}\label{table1}
\end{subtable}

\bigskip
\begin{subtable}{1\textwidth}
\centering
   \begin{tabular}{|l|c|c|c|c|c|}
      \hline
      {\(\;\;\;\;\;\;\;\;\;\;\;\;\;\;\;\;n\)} & {\(10^4\)} & {\(10^5\)} & {\(10^6\)} & {\(10^7\)} & {\(10^8\)}\\ 
      \hline
      Recursive upper bound & 102.198 & 280.508 & 886.491 & 2289.000 & 6700.205\\ 
      \hline
      Upper bound from \eqref{eq_ub_ring} & 132.987 & 332.071 & 1031.794 & 2516.117 & 7245.866\\
      \hline
   \end{tabular}
   \caption{Version age of a single node when $\alpha = 0.1$.}\label{table2}
\end{subtable}

\bigskip
\begin{subtable}{1\textwidth}
\centering
   \begin{tabular}{|l|c|c|c|c|c|}
      \hline
      {\(\;\;\;\;\;\;\;\;\;\;\;\;\;\;\;\;n\)} & {\(10^4\)} & {\(10^5\)} & {\(10^6\)} & {\(10^7\)} & {\(10^8\)}\\ 
      \hline
      Recursive upper bound & 67.926 & 178.614 & 445.018 & 1101.664 & 2805.389\\ 
      \hline
      Upper bound from \eqref{eq_ub_ring} & 82.213 & 197.497 & 469.331 & 1133.706 & 2851.794\\
      \hline
   \end{tabular}
   \caption{Version age of a single node when $\alpha = 0.2$.}\label{table3}
\end{subtable}

\bigskip
\begin{subtable}{1\textwidth}
\centering
   \begin{tabular}{|l|c|c|c|c|c|}
      \hline
      {\(\;\;\;\;\;\;\;\;\;\;\;\;\;\;\;\;n\)} & {\(10^4\)} & {\(10^5\)} & {\(10^6\)} & {\(10^7\)} & {\(10^8\)}\\ 
      \hline
      Recursive upper bound & 46.186 & 101.728 & 224.936 & 503.411 & 1121.325\\ 
      \hline
      Upper bound from \eqref{eq_ub_ring} & 57.45 & 113.806 & 237.864 & 517.252 & 1136.084\\
      \hline
   \end{tabular}
   \caption{Version age of a single node when $\alpha = 0.3$.}\label{table4}
\end{subtable}

\caption{We enumerate the upper bounds for the version age in the generalized ring network when $f(n) = n^\alpha$, where $\alpha = 0, 0.1, 0.2, 0.3$.} \label{tab:three_tables}
\end{table}

\subsection{Numerical Results for the Generalized Ring Network}
Next, we perform simulations to compare our upper bounds with the simulated version age of a single node for the generalized ring network. The upper bound for the version age of the generalized ring network depends on the number of nodes $n$, number of connections $2f(n)$, and the information flow rates $\lambda_e$ and $\lambda$. We again choose $\lambda_e = \lambda = 1$. 

We plot the version age for $f(n) = n^\alpha$ for $\alpha$ ranging from $0.4$ to $0.9$ with increments of $0.1$. The number of nodes varies from $1000$ to $5000$ with a step size of $1000$. We plot the results in Fig.~\ref{fig_ring_no_log} and Fig.~\ref{fig_ring_log}. In Fig.~\ref{fig_ring_no_log}, the x-axis is in a linear scale. We see that the version age decreases as $\alpha$ increases, which is consistent with the theoretical upper bound result. In Fig.~\ref{fig_ring_log}, the x-axis is in log scale. We see that the version age plots in Fig.~\ref{fig_ring_log} appear as straight lines, thus showing that they have approximate $\log$ scaling for low values of $n$, satisfying Remark~\ref{remark_generalized_rings}.

We do not perform simulations for lower values of $\alpha$ because the functions $f(n)$ grow very slowly. For small values of $\alpha$, the value of $f(n)$ might be constant, even if the number of nodes increases.  Hence, for $\alpha$ between $0$ and $0.3$, we calculate the upper bound that we obtain from the recursive equations in \eqref{eq_X_rub}, \eqref{eq_Y_rub} and \eqref{eq_Z_rub}, and compare them to the upper bound obtained in \eqref{eq_ub_ring}. These results are summarized in Tables~\ref{table1}, \ref{table2}, \ref{table3} and \ref{table4}.

\subsection{Numerical Results for the Unit Hypercube}
\begin{figure}[t]
        \centering
        \includegraphics[width = 0.7\textwidth]{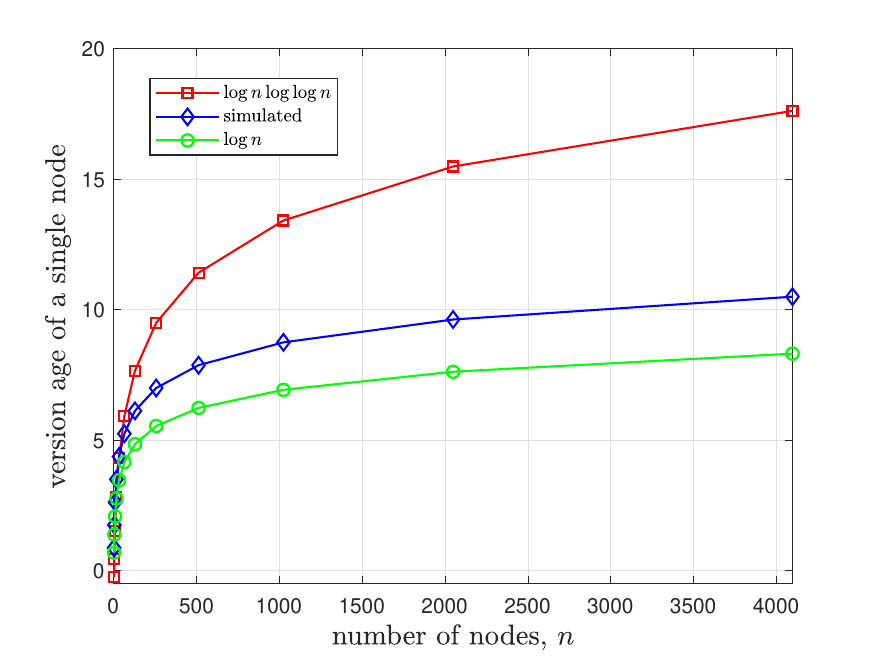}
        \caption{Version age of a single node versus the number of nodes in a unit hypercube.}
        \label{fig_unit_hypercube}    
\end{figure}

We now simulate the unit hypercube network. We choose $\lambda_e = \lambda = 1$. We choose $m$ ranging from $1$ to $12$ with step size of $1$. We plot the results in Fig.~\ref{fig_unit_hypercube} along with $\log{n}\log{\log{n}}$ and $\log{n}$. We notice that, for small values of $n$, $\log{n}\log{\log{n}}$ is negative. However, for larger values of $n$, the upper bound is satisfied. Moreover, $\log{n}$ is a lower bound for $v_1$ for all values of $n$ that we choose. We note that the simulated version age follows the $\log{n}$ curve more closely than the $\log{n}\log{\log{n}}$ curve.

\subsection{Numerical Results for the General Hypercube}
\begin{figure}[t]
    \centering
    \begin{subfigure}{0.48\textwidth}
        \centering
        \includegraphics[width = \textwidth]{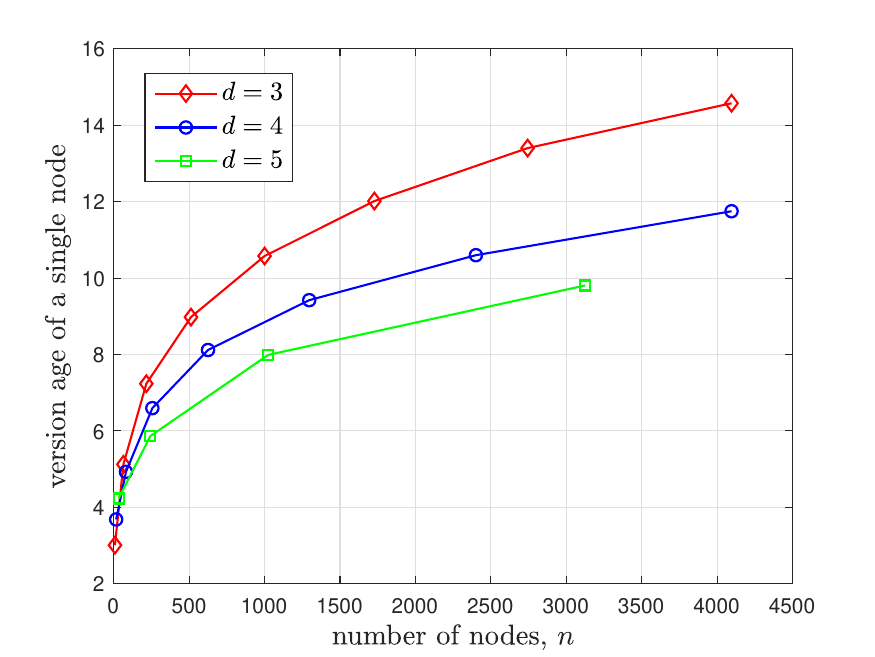}
        \caption{}
        \label{fig_d_dim_grid}
    \end{subfigure}
    \hfill
    \begin{subfigure}{0.48\textwidth}
        \centering
        \includegraphics[width = \textwidth]{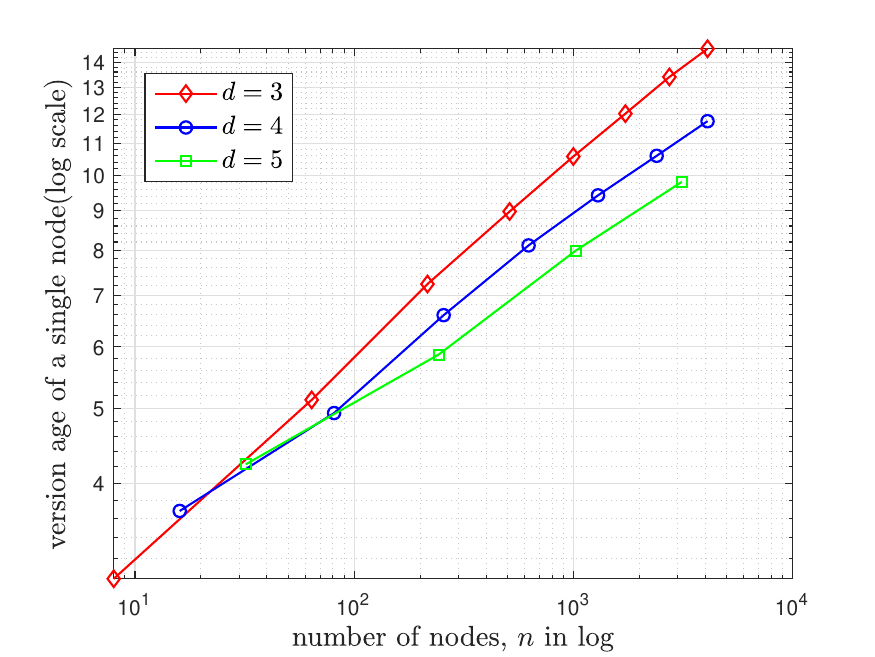}
        \caption{}
        \label{fig_d_dim_grid_log}
    \end{subfigure}
    \caption{Variation of the version age of a single node in a $d$-dimensional grid where $d=3,4,5$. The left figure is in linear scale, and the right figure is in log scale.}
\end{figure}

Finally, we simulate the general hypercube network. Again, we choose $\lambda_e = \lambda = 1$. We simulate the $d$-dimensional grid for $d=3, 4, 5$. For $d=3$, we simulate for the range $m=2$ to $m=16$ with step size $2$. For $d=4$, we simulate for the range $m=2$ to $m=8$ with step size $1$. For $d=5$, we simulate for the range $m=2$ to $m=5$ with step size $1$. We plot the results in Fig.~\ref{fig_d_dim_grid}, where the plot is on a linear scale, and in Fig.~\ref{fig_d_dim_grid_log}, where the plot is in the log scale. We see from Fig.~\ref{fig_d_dim_grid} that the version age of a single node is a growing function of the number of nodes in the network, and the version age decreases as $d$ increases, for a given number of nodes. We also plot the values in log-log scale since the version age scaling is $O(n^{\frac{1}{d+1}})$, the log-log plot will be linear with slope $\frac{1}{d+1}$. In fact, the slopes for $d=3, 4, 5$ are $0.2478, 0.2072$ and $0.1836$, respectively. Hence, for $d=3$ and $4$, the slopes are consistent with the exact slope. However, for $d=5$, the slope is a little higher than the expected $0.16667$. This is because the number of nodes in the simulations is not large enough to get the exact scaling.

\section{Conclusion}\label{sec11}
In this work, we provided a new technique to find a tight upper bound for the version age of a connected subset of nodes in a gossiping network by modifying the recursive equations found in \cite{yates21gossip}. We found these bounds in terms of, first, the number of neighbors of the set, and then the number of neighboring edges of the set. We used this technique to analyze various graph models that are used extensively in the literature, including the two-dimensional grid graph, generalized rings with varying number of neighbors, unit hypercube graphs with varying dimension and fixed side length, and hypercube graphs in fixed dimension with varying side length. In this analysis, we found the version age of a single node in each network. We extended the results of the previously analyzed graphs, the fully-connected network and the bi-directional ring network and verified the bounds obtained in both cases. Moreover, we obtained a sequence of graphs that go from the most disconnected case, which is the bi-directional ring network to the most connected case, which is the fully-connected network, through the generalized ring network analysis. In this way, we were able to get insights into the graph properties which version age depends on. 

\bibliographystyle{unsrt}
\bibliography{refs}

\end{document}